\documentclass[12pt,a4paper]{article}

\usepackage{amsfonts,amssymb,latexsym,amsmath,amscd,tikz}
\usepackage{youngtab}
\usepackage{mathrsfs}
\usepackage{times}
\usepackage{anysize}

\usepackage{youngtab}
\usepackage{rotfloat}
\usepackage{rotating}
\usepackage{tikz}
\usepackage{shadow}
\usepackage{fancybox}
\usepackage{xcolor}
\usepackage{float}

\usepackage{times}
\usepackage{jheppub}
\makeatletter
\def\@fpheader{\relax}
\makeatother
\numberwithin{figure}{section}
\usepackage{graphicx}

\usepackage{framed}
\definecolor{shadecolor}{rgb}{0.90,0.90,0.90}

\usepackage{tikz}
\usetikzlibrary{automata}
\usetikzlibrary{arrows}
\usetikzlibrary{calc}
\usetikzlibrary{decorations.markings}
\usetikzlibrary{decorations.pathreplacing}
\usetikzlibrary{intersections}
\usetikzlibrary{positioning}
\usetikzlibrary{topaths}
\usetikzlibrary{shapes.geometric}
\usetikzlibrary{shapes.misc}
\tikzset{->-/.style = {
    decoration = {markings, mark = at position #1 with {\arrow{>}}},
    postaction = {decorate}}}
\tikzset{color-group/.style = {
    shape = circle,
    minimum size = 2.5ex,
    inner sep = .5ex,
    draw}}
\tikzset{flavor-group/.style = {
    shape = rectangle,
    minimum size = 2.5ex,
    inner sep = .5ex,
    draw}}
\tikzset{cf-group/.style = {
    shape = rounded rectangle,
    rounded rectangle right arc = none,
    draw}}
\tikzset{fc-group/.style = {
    shape = rounded rectangle,
    rounded rectangle left arc = none,
    draw}}
\tikzset{cross/.style={minimum width=1pt, path picture={
      \draw[black, very thick]
               (path picture bounding box.south east)
            -- (path picture bounding box.north west)
               (path picture bounding box.south west)
            -- (path picture bounding box.north east);
          }}}

\def\bea{\begin{eqnarray}}
\def\eea{\end{eqnarray}}
\def\be{\begin{equation}}
\def\ee{\end{equation}}
\def\ba{\begin{align}}
\def\ea{\end{align}}

\newtheorem{lemma}{Lemma}[section]
\newtheorem{conjecture}[lemma]{Conjecture} 
 
\newtheorem{theorem}[lemma]{Theorem} 
\newtheorem{definition}[lemma]{Definition}

\newcommand{\C}{\mathbb{C}}
\newcommand{\CC}{\mathbb{C}}

\newcommand{\Hh}{\mathbf{H}}
\newcommand{\ZZ}{\mathbb{Z}}
\newcommand{\End}{\mathrm{End}}

\newcommand{\bC}{\ensuremath{\mathbb{C}}}

\newcommand{\bK}{\ensuremath{\mathbb{K}}}
\newcommand{\bL}{\ensuremath{\mathbb{L}}}

\newcommand{\bP}{\ensuremath{\mathbb{P}}}

\newcommand{\bR}{\ensuremath{\mathbb{R}}}

\newcommand{\bZ}{\ensuremath{\mathbb{Z}}}

\newcommand{\scC}{\ensuremath{\mathscr{C}}}

\newcommand{\scH}{\ensuremath{\mathscr{H}}}

\newcommand{\scL}{\ensuremath{\mathscr{L}}}

\newcommand{\scP}{\ensuremath{\mathscr{P}}}

\newcommand{\fraka}{\ensuremath{\mathfrak{a}}}

\newcommand{\frakg}{\ensuremath{\mathfrak{g}}}

\newcommand{\frakq}{\ensuremath{\mathfrak{q}}}

\newcommand{\frakt}{\ensuremath{\mathfrak{t}}}

\newcommand{\fraksl}{\ensuremath{\mathfrak{sl}}}
\newcommand{\frakso}{\ensuremath{\mathfrak{so}}}
\newcommand{\fraksp}{\ensuremath{\mathfrak{sp}}}

\newcommand{\cH}{\mathcal{H}}

\newcommand{\cL}{\mathcal{L}}
\newcommand{\cM}{\mathcal{M}}
\newcommand{\cN}{\mathcal{N}}
\newcommand{\cO}{\mathcal{O}}

\newcommand{\GL}{\mathrm{GL}}
\newcommand{\SU}{\mathrm{SU}}

\newcommand{\SL}{\mathrm{SL}}

\newcommand{\U}{\mathrm{U}}

\newcommand{\Spin}{\mathrm{Spin}}

\newcommand{\Tr}{\mbox{Tr}}

\def\SU{\mathrm{SU}}
\def\SL{\mathrm{SL}}

\def\U{\mathrm{U}}
\def\bC{\mathbb{C}}

\def\bZ{\mathbb{Z}}

\DeclareMathOperator{\Hilb}{Hilb}
\DeclareMathOperator{\Spec}{Spec}

\DeclareMathOperator{\ch}{ch}

\DeclareMathOperator{\MF}{MF}
\DeclareMathOperator{\HHH}{HH}

\newtheorem{fact_}{Fact}[section]
{\begin{fact_}\begin{shaded}}%
{\end{shaded}\end{fact_}}

\newtheorem{fact?_}[fact_]{Fact?}
\newenvironment{fact?}%
{\begin{fact?_}\begin{shaded}}%
{\end{shaded}\end{fact?_}}

\let\oldtext\text
\def\text#1{\oldtext{\upshape #1}}

\newcommand{\unknot}{{\raisebox{-.11cm}{\includegraphics[width=.37cm]{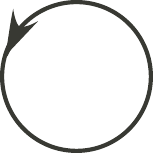}}}\,}

\def\wt{\widetilde}

\def\bar{\overline}

\begin{document}
\Yboxdim4pt

\preprint{}
\title{Lectures on knot homology}
\author{Satoshi Nawata${}^1$}
\author{Alexei Oblomkov${}^2$}
\emailAdd{snawata@gmail.com}
\emailAdd{oblomkov@math.umass.edu}
\affiliation[1]{Faculty of Physics, University of Warsaw,  ul. Pasteura 5,  02-093, Warsaw, Poland\\
Max-Planck-Institut f\"ur Mathematik, Vivatsgasse 7, D-53111 Bonn, Germany
}
\affiliation[2]{Department of Mathematics,
University of Massachusetts at Amherst
}
\abstract{We  provide various formulations of knot homology that are predicted by string dualities. In addition, we also explain the rich algebraic structure of knot homology which can be understood in terms of geometric representation theory in these formulations. 

These notes are based on lectures in the workshop ``Physics and Mathematics of Link Homology'' at Centre de Recherches Math\'ematiques, Universit\'e de Montr\'eal.
}

\maketitle

\section{Introduction}
For past decades, we have witnessed the fruitful interplay of mathematics and physics involving knot theory. In particular, the formulation of Jones polynomial \cite{Jon} in terms of Chern-Simons theory \cite{CSJones} has led to intensive study of quantum knot invariants. These quantum invariants have been also realized in the context of string theory \cite{OV}. Around the same time,
Khovanov \cite{Khovanov:1999qla} constructed the bi-graded homology which is itself a knot invariant and its graded Euler characteristics is the Jones polynomial. The categorifications of quantum knot invariants \cite{Khovanov:1999qla,KR1,KR2} initiated by Khovanov reveal a truly groundbreaking perspective to knot theory not only because they are more powerful than quantum invariants but also because they are functorial. These homological invariants have been interpreted in many duality frames in string theory, which provide various aspects of knot homology. Hence, one of the main aims in these notes is to account for many facets of knot homology.

The various dualities in string theory provides different approaches to knot homology. Nevertheless, in any vantage point we take, the main idea in physics is that the knot homology can be interpreted as a space of BPS states \cite{GSV}:
\bea\nonumber
\cH_{\textrm{knot}}\cong\cH_{\textrm{BPS}}~.
\eea
The advantage of categorification program stems from the fact that knot homology encodes more information than quantum knot invariants. As various physical viewpoints predict algebraic structure of BPS states,  richer structure becomes manifest only at the categorified level. Because BPS states are indeed realized as cohomology groups of some moduli spaces, the structure of BPS states appears as geometric realization of a certain representation of an algebra in mathematics. Therefore, the knot homology is naturally  connected to geometric representation theory:
\begin{center}
 \shabox{\parbox{.85\hsize}{Algebraic structure of BPS states \ \ \  $ \phantom{\int} \Leftrightarrow \phantom{\int} $ \ \ \  Geometric representation theory}}
\end{center}
Through the recent developments \cite{OS,GOR, GORS} in mathematics, it turns out that the underlying structure of torus knot homology is governed by the \emph{rational Cherednik algebra}. This discovery has led to the precise formulations of the geometric transition and the geometric engineering in mathematics. Moreover, in some special settings, these string dualities have been proven based on geometric representation theory of the rational Cherednik algebra. Therefore, the other objective of the notes is to describe the structural properties of knot homology and their connections to geometric representation theory.

\subsection{Physical setups}

All physical approaches to knot homology essentially originate in the two physical frameworks, ``deformed conifold'' and ``resolved conifold'', related by the {\it geometric transition}  \cite{Mtop1,Mtop2,gauge-gravity}. On the ``deformed conifold'' side, the physical setups are as follows;

\noindent
\begin{small}
\begin{minipage}[b]{7.5cm}
\be\label{deformed}
\begin{matrix}
{\mbox{\rm space-time:}} & \quad & \bR_t & \times & TN_4 & \times  &T^* S^3 \\
{\mbox{\rm $N$ M5-branes:}} & \quad & \bR_t &\times& D & \times &  S^3  \\
{\mbox{\rm M5'-brane:}} & \quad&  \bR_t &\times& D & \times & \cL_K  
\end{matrix}\nonumber
\ee
\end{minipage}
\begin{minipage}[b]{7.5cm}
\be\label{deformed}
\begin{matrix}
{\mbox{\rm space-time:}} & \quad & \bR_t & \times & TN_4 & \times  &T^* S^3 \\
{\mbox{\rm $N$ M5-branes:}} & \quad & \bR_t &\times& D & \times &  S^3  \\
{\mbox{\rm M2'-brane:}} & \quad&  \bR_t &\times& \textrm{pt} & \times & C_K  
\end{matrix}
\ee
\end{minipage}
\end{small}
\vspace{.2cm}

\noindent where $TN_4 \cong \bR^4_{\epsilon_1,\epsilon_2}$ is the Taub-NUT space,
$D \cong \bR^2_{\epsilon_1}$ is the two-dimensional ``cigar'' (holomorphic Lagrangian submanifold) in the Taub-NUT space. The $N$ M5-branes wrap the zero section (the special Lagrangian subvariety) of the cotangent bundle $T^*S^3$ of a 3-sphere, realizing Chern-Simons theory on $S^3$ \cite{CSstring}. A knot is created either by a spectator M5'-brane (left) or by a spectator M2'-brane (right). The spectator M5'-brane sits on another Lagrangian subvariety, which is the co-normal bundle $\cL_K \subset T^*S^3$ to a knot $K \subset S^3$ where the knot $K$ is realized as the intersection of the two stacks of M5-branes $K = S^3 \cap \cL_K$ \cite{OV}. On the other hand, the spectator M2'-brane on $C_K$ attaches to $N$ M5-branes at a knot $K$, extending to the fiber direction \cite{DSV,DHS,fiveknots}.

At large $N$, the geometry undergoes the transition where $S^3$ shrinks and $S^2=\bP^1$ is blown up\footnote{As $S^3$ shrinks, in order for the spectator brane to avoid the singularity, the Lagrangian submanifold $\cL_K$ is lifted to the fiber direction and it no longer touches  $S^3$. We refer the reader to \cite{DSV} for detailed treatment.}:

\noindent
\begin{minipage}[b]{7.5cm}
\be
\begin{matrix}
{\mbox{\rm space-time:}} & \quad & \bR_t & \times & TN_4 & \times & X \\
{\mbox{\rm M5'-brane:}} & \quad & \bR_t & \times & D & \times & \cL_K
\end{matrix}\nonumber
\ee
\end{minipage}
\begin{minipage}[b]{7.5cm}
\be\label{resolved}
\begin{matrix}
{\mbox{\rm space-time:}} & \quad & \bR_t & \times & TN_4 & \times & X \\
{\mbox{\rm M2'-brane:}} & \quad & \bR_t & \times & \textrm{pt} & \times & C_K  \\
\end{matrix}
\ee\end{minipage}
\vspace{.2cm}

\noindent Here $X=\mathcal{O}(-1)\oplus \mathcal{O}(-1) \to\bP^1$ is the resolved conifold. Through the geometric transition, the stack of $N$ M5-branes turns into the flux supported on $\bP^1$ and the volume (K\"ahler parameter) of  $\bP^1$ is expressed by the $a$ variable. In the contrast, the spectator brane remains and in particular the spectator M5'-brane wraps a Lagrangian submanifold of $X$ associated to the knot $K$. In the left setting, the BPS states are indeed represented by M2-branes wrapping a 2-cycle $\beta\in H_2(X,\cL_K)$ and ending on the M5'-brane.  All spaces of BPS states receive the equivariant action of $\Spin(2)_L \times {\Spin}(2)_R$ that is the symmetry group of on the tangent and normal bundle of $D \subset TN_4$. Thus, we count the BPS states with the following weight:
\bea
a\textrm{-grading}&=& 2\beta\in H_2(X,\cL_K) \cr
q\textrm{-grading}&=& 2j_L +2n \cr
t\textrm{-grading}&=&  2j_R\nonumber
\eea
where $(j_L,j_R)$ is the spin of  $\Spin(2)_L \times \Spin(2)_R$ and $n$ is a Kaluza-Klein mode of M2-brane (D0-brane charge). In the right setting, the spectator M2'-brane bound together with a closed M2-brane wrapping $\bP^1$ forms the BPS states  where the $a$-gradings are labelled by how many times the M2-brane wraps $\bP^1$.

Hence, as we have seen, the ``deformed conifold'' provides ways to approach to doubly-graded $\fraksl(N)$ homology for fixed $N$
whereas the ``resolved conifold'' leads to triply-graded HOMFLY homology:
 \begin{center}
 \shadowbox{\begin{minipage}{.88\linewidth}
\be\nonumber
\begin{array}{rcl}
\text{deformed conifold} & \quad \phantom{\int} \Leftrightarrow \phantom{\int} \quad & \textrm{doubly-graded $\fraksl(N)$ homology}  \\
\text{resolved conifold} & \quad \phantom{\int} \Leftrightarrow \phantom{\int} \quad & \textrm{triply-graded HOMFLY homology}
\end{array}
\ee
\end{minipage}}
\end{center}
\vspace{-.3cm}

Moreover,  the vantage point of the Taub-NUT space $\bR_t\times TN_4$  in the left of \eqref{resolved} leads to $\U(1)$ gauge theory on  $\bR_t\times TN_4$ with codimension two defect supported on $\bR_t\times D$ associated to the knot $K$, which is called \emph{geometric engineering} \cite{GeomEng}. From this point of view, one can express the counting of BPS states in terms of equivariant instanton counting on $TN_4$ in the presence of codimension two defect where $a$ is the Coulomb branch parameter and, $(q,t)$ are the $\bC^*\times \bC^*$-equivariant parameters.

Categorifications are closely related to refinements in physics. In \cite{AS}, ``refinement'' of Chern-Simons theory has been proposed by using a extra $\U(1)$ symmetry where the main ingredients are the Macdonald deformation of modular $S$ and $T$ matrices \cite{Kirillov}. These provide a direct way to compute refined Chern-Simons invariants of torus knots whose stable limits at large $N$ are conjectured to be equal to Poincar\'e polynomials of HOMFLY homology in the case of rectangular Young diagrams.

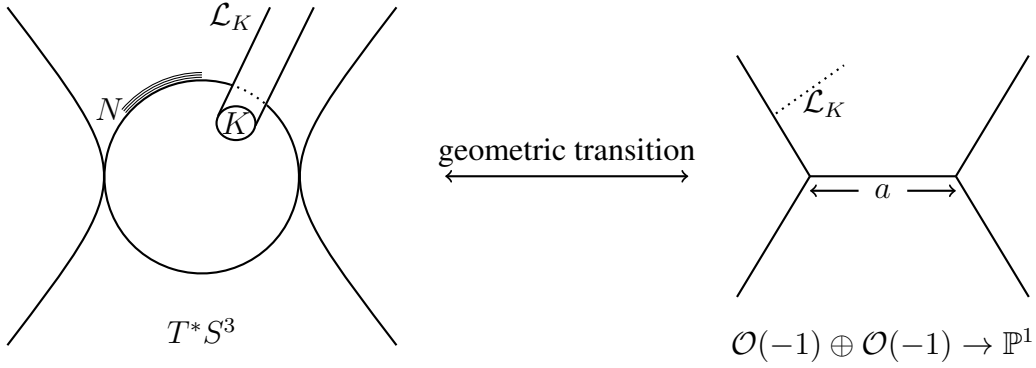
\begin{figure}[h]\centering
\begin{tikzpicture}[scale=0.64]
[place/.style={circle,draw=blue!50,fill=blue!20,thick},
transition/.style={rectangle,draw=black!50,fill=black!20,thick}]
\draw[thick] (.7,1.1) ellipse (.4cm and .35cm);
\draw[thick] (4,-3.5) .. controls (1.35,0)  .. (4,3.5);
\draw[thick] (-4,-3.5) .. controls (-1.35,0)  .. (-4,3.5);
\draw[thick] (.33,1.25) -- (1.4,3.5);
\draw[thick] (1.1,1.05) -- (2.3,3.5);

\draw[thick] (.58,1.9)
        arc [x radius = 2cm, y radius =  2cm, start angle = 73, end angle = 408];
        \draw[thick,dotted] (.58,1.9)
        arc [x radius = 2cm, y radius =  2cm, start angle = 73, end angle = 48];
\draw[thin] (0,2.15)
        arc [x radius = 2.15cm, y radius =  2.15cm, start angle = 90, end angle = 140];
\draw[thin] (0,2.1)
        arc [x radius = 2.1cm, y radius =  2.1cm, start angle = 90, end angle = 140];
\draw[thin] (0,2.05)
        arc [x radius = 2.05cm, y radius =  2.05cm, start angle = 90, end angle = 140];
        
  \node () at (0,-3.2)  {$T^*S^3$}  ;
    \node () at (.7,1.1)  {$K$}  ;
        \node () at (.6,3.3)  {$\mathcal{L}_K$}  ;
  \node () at (-1.9,1.4)  {$N$}  ;

  \draw[thick,<->]  (5,0) to (10,0);
    \node () at (7.5,.5)  {geometric transition}  ;

  \draw[thick]  (11,2.5) to (12.5,0);
  \draw[thick]  (11,-2.5) to (12.5,0);
  \draw[thick] (12.5,0) to (15.5,0);
  \draw[thick]  (17,2.5) to (15.5,0);
  \draw[thick]  (17,-2.5) to (15.5,0);
  \draw[thick,dotted]  (11.8,1.3) to (13.2,2.3);
    \draw[thick,->] (14.5,-.3) to (15.5,-.3);
     \draw[thick,->] (13.5,-.3) to (12.5,-.3);

    \node () at (12.8,1.5)  {$\mathcal{L}_K$}  ;
   \node () at (14,-.3)  {$a$}  ;
    \node () at (14,-3.5)  {$\mathcal{O}(-1)\oplus \mathcal{O}(-1) \to \bP^1$}  ;
\end{tikzpicture}
\caption{Schematic illustration of the geometric transition with the spectator M5'-brane.}
\end{figure}

\vspace{-.2cm}
\subsection{Mathematical formulations}

In recent years, mathematics of the ``resolved conifold'' side has been investigated  in the case of torus knots. Motivated by  conjectures from \cite{OS}, the geometric transition with the spectator M2'-brane has been proposed in  \cite{DSV}. In particular, for a link $L_{C,(0,0)}$ created as an intersection of a planar curve $C:=\{E(x,y)=0\}$ with the 3-sphere $|x|^2+|y|^2=r$, the configuration of the spectator M2'-brane in \eqref{resolved} turns out to be exactly $E(x,y)=0$ where we identify the fiber of the resolved conifold $X$ with complex coordinates $(x, y)$. Specifically, for an $(m,n)$ torus knot $T_{m,n}$, the curve $C_{m,n}$ is represented by $E(x,y)=x^m-y^n$.

Then, the moduli space of M2-M2' bound states is isomorphic to the Hilbert scheme of points on $C$. In fact, the reduction to type IIA theory provide D2-D2'-D0 bound states that describe the vortex configurations and the moduli space of the vortex configurations is indeed  the Hilbert scheme of points on $C$. Therefore, the moduli space is expressed in terms of the geometry of the plane curve singularity without involving the resolved conifold $X$. Now, let us state the precise formulation on the relation between HOMFLY homology and Hilbert schemes of the planar curve $C$.
On $C$, we consider the moduli space $C_0^{[n]}$ parameterizing 
ideals in the local ring of the point $(0,0)$ of colength $n$,
where $\mathcal{E}$ is a torsion free sheaf. As shown in \cite{PT1}, 
these spaces are isomorphic to the moduli spaces of pairs $s:\cO_C \to \mathcal{E}$, 
and $\dim \mathcal{E}/s\cO_C = n$,  for Gorenstein (and in particular planar) 
curves $C$. Then, we can define the nested Hilbert schemes:
$C_0^{[n,n+m]}\subset C^{[n]}_0\times C_0^{[n+m]}$ consisting of pairs of ideals $(I,J)$ inside the local ring $\mathcal{O}_0$ such that $\mathfrak{m}I\subset J\subset I$ where 
$\mathfrak{m}$ is the maximal ideal of $\mathcal{O}_0$. Based on the conjecture on the HOMFLY polynomial \cite{OS} (now theorem of \cite{Mau}), the relation between HOMFLY homology and the nested Hilbert schemes is conjectured in the following way. We shall account for more details in \S\ref{subsec:geometry}.
\begin{conjecture}\label{planar-homology}\text{\cite{ORS}}
Let C the germ of a plane curve singularity. With notation as above, the Poincar\'e polynomials of the (triply graded) HOMFLY homology of the link of the singularity is
\be
\overline\scP_{\yng(1)}(L_{C,(0,0)};a,q,t)=(a/q)^{\mu-1}\sum_{m,n} a^{2m} q^{2n} t^{m^2}  P_t( C^{[n,n+m]}),
\ee
where $\mu = \dim \bC[[x,y]]/(\partial_x E, \partial_y E)$
is the Milnor number of the singular point and $P_t$ is the virtual Poincar\'e polynomial\footnote{The virtual Poincar\'e polynomial is uniquely defined by the following properties.
For smooth projective $X$, $P_t(X)$ is the usual Poincar\'e polynomial and $P_t(Z)=P_t(X)+P_t(Y)$ if $Z= X\sqcup Y$  and $X$, $Y$ are algebraic.}.
\end{conjecture}

By the reduction of the right configuration in \eqref{resolved} on the circle of the cigar in $TN_4$, one obtains type IIA theory with D6-D2-D2'-D0 brane configurations. In this setup, the BPS states are ``refined'' D6-D2-D2'-D0 bound states that are mathematically ``motivic'' Donaldon-Thomas invariants in the resolved conifold. The mathematical construction of these invariants is described by stable pairs. Given a singular plane 
curve $C^\circ$ in a fiber of the projection $X\to\bP^1$, there is a 
natural moduli space $\mathcal{P}(X,C,p,n,r)$ of $C^\circ$-framed stable pairs on $X$. 
These are pairs $s:\cO_X \to\mathcal{G}$ on $X$ 
where $\mathcal{G}$ is topologically supported on the union of $C^\circ$ 
with the zero section $\bP^1\subset X$, and has multiplicity one along $C^\circ$. The main claim in \cite{DHS} is that colored HOMFLY homology can be expressed by stable pairs subject to the framing condition. Furthermore, such 
moduli spaces are related to the nested Hilbert schemes 
above by a variation 
of stability condition, which will be explained  in \S\ref{subsec:colored pairs}.

Now let us specialize to the case of torus knots $C_{m,n}:=\{x^m=y^n\}$.
It was observed in \cite{OY1} that the space of the cohomology group $\oplus_\ell H^*(C_{m,n}^{[\ell]},\mathbb{Q})$ of the Hilbert schemes has a structure of a module of the spherical rational Cherednik algebra $e\overline \Hh_{\frac{m}{n}}(S_n)e$.
In fact, it is the finite-dimensional irreducible module denoted by $e\overline L_{\frac{m}{n}}$ \cite{BEG3}.
Thus, via Conjecture \ref{planar-homology}, we obtain a representation theoretic interpretation of HOMFLY homology of torus knots (see \S\ref{subsec:torus ratDAHA} for more details).

It was known previously \cite{GS,GS2} that to a large class of modules over the rational Cherednik algebra $\overline\Hh_{\frac{m}{n}}(S_n)$ one can attach a canonical $\CC^*\times\CC^*$-equivariant sheaf on 
the Hilbert scheme $\Hilb_n(\CC^2)$ of points on $\CC^2$. Combining the intuition of \cite{GS,GS2}, the authors of \cite{GORS} obtained the conjectural description of HOMFLY homology of torus knots in terms
of the sheaves on the Hilbert scheme $\Hilb_n(\CC^2)$. Below we state a simplified version of the conjecture:

\begin{conjecture}\text{\cite{ORS}} There is a $\CC^*\times\CC^*$-equivariant sheaf $F_{m,n}$ on $\Hilb_n(\CC^2)$ such that 
$$\overline\scP_{\yng(1)}(T_{m,n};a=0,q,t)=\chi_{\frakq,\frakt}(F_{m,n})~,$$
where $\chi_{\frakq,\frakt}$ is the equivariant Euler characteristics. The identity requires the change of variables \eqref{cov}.
\end{conjecture}

Actually, this conjecture paves the way to mathematically formulate the geometric engineering \cite{GeomEng} in the case of torus knots. Presumably, the  sheaf $F_{m,n}$ describes cohomology groups of the moduli space of $\U(1)$ instanton in the presence of codimension two defect associated to the torus knot $T_{m,n}$. In fact, the construction of the sheaf $F_{m,n}$ is explicitly known for the case $m=nk+1$ in the work of \cite{GS2}. The conjectural description of the sheaf $F_{m,n}$ for general $m,n$ is obtained in \cite{GN}. To formulate the geometric engineering, the authors of \cite{GN} make use of Cherednik's interpretation of the superpolynomial for $T_{m,n}$ in terms of the double affine Hecke algebra (DAHA). Motivated by refined Chern-Simons theory \cite{AS}, DAHA-Jones polynomials of torus knots have been defined by using $\textrm{PSL}(2,\bZ)$ transformations in DAHA. They are the same as refined Chern-Simons invariants 
and their stable limits at large $N$ are called DAHA-superpolynomials (see \S \ref{sec:DAHA} for more details).
We devote \S\ref{subsec:torus Hilb}  to the discussion of the geometric ideas behind the construction of $F_{m,n}$.  Below we state an explicit formula for the DAHA-superpolynomial which follows by localization
from the description of $F_{m,n}$. Indeed the formula takes the form of the Nekrasov partition functions.

\begin{theorem}\text{\cite{GN}}
\label{thm:main}
The unreduced DAHA-superpolynomial $\overline\scP^{\textrm{DAHA}}_{\yng(1)}(T_{m,n};\mathfrak{a},\mathfrak{q},\mathfrak{t})$, defined in \cite{Ch}, is given by
\bea
\label{eqn:for}
&&\overline\scP_{\yng(1)}^{\textrm{DAHA}}(T_{m,n};\mathfrak{a},\mathfrak{q},\mathfrak{t})\\
&& =\sum_{\mu \vdash n} \frac {\widetilde{\gamma}^n}{\widetilde{g}_\mu} \sum^{\textrm{SYT}}_{\text{of shape }\mu}  \frac {\prod_{i=1}^{n} \chi_i^{S_{\frac{m}{n}}(i)} (1-\mathfrak{a}\chi_i)(\mathfrak{q} \chi_i - \mathfrak{t})}{\left(1-\frac {\mathfrak{q}\chi_{2}}{\mathfrak{t}\chi_{1}}\right) \ldots  \left(1-\frac {\mathfrak{q}\chi_{n}}{\mathfrak{t}\chi_{n-1}}\right)}\prod_{1\leq i < j\leq n} \frac {(\chi_j-\mathfrak{q}\chi_i)(\mathfrak{t}\chi_j-\chi_i)}{(\chi_j-\chi_i)(\mathfrak{t}\chi_j-\mathfrak{q}\chi_i)}~,\nonumber
\eea
where the sum is over all standard Young tableaux of size $n$, and $\chi_i$ denotes the $(\mathfrak{q},\mathfrak{t}^{-1})$-weight of the box labeled by $i$ in the tableau. The constants in the above relation are given by
\begin{equation}
\label{def:smn}
S_{\frac{m}{n}}(i)=\left\lfloor \frac {im}n\right\rfloor - \left\lfloor \frac {(i-1)m}n \right\rfloor, \qquad \qquad \widetilde{\gamma}=\frac {(\mathfrak{t}-1)(\mathfrak{q}-1)}{(\mathfrak{q}-\mathfrak{t})}~,
\end{equation}
and
$$
\widetilde{g}_\mu =\prod_{\square \in \lambda}(1-\mathfrak{q}^{a(\square)}\mathfrak{t}^{l(\square)+1}) \prod_{\square \in \lambda}(1-\mathfrak{q}^{-a(\square)-1}\mathfrak{t}^{-l(\square)})~.
$$
We denote arm-length and leg-length of a box in a Young diagram  by $a(\square)$  and $l(\square)$ .
\end{theorem}

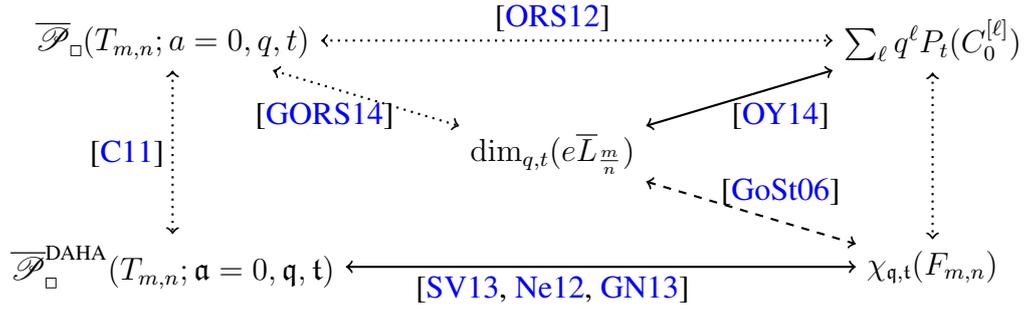
\begin{figure}[h]
\centering
\begin{tikzpicture}
   \node (A) at (0,0) {$\overline\scP_{\yng(1)}(T_{m,n};a=0,q,t)$} ;   
  \node (B)at (0,-3)  {$\overline\scP_{\yng(1)}^{\textrm{DAHA}}(T_{m,n};\fraka=0,\frakq,\frakt)$} ;   
  \node(C) at (10,0)   {$\sum_{\ell}q^\ell {P}_t(C_0^{[\ell]})$};
  \node (D) at (5,-1.5)   {$\dim_{q,t}(e\overline L_{\frac{m}{n}})$};
    \node (E) at (10,-3)   {$\chi_{\frakq,\frakt}(F_{m,n})$};

  \draw[thick,dotted,<->]  (A) to (B);
   \draw[thick,dotted,<->]  (A) to (C);
   \draw[thick,dotted,<->]  (A) to (D);
    \draw[thick,dotted,<->]  (C) to (E);
    \draw[thick,<->]  (B) to (E);    
    \draw[thick,dashed,<->]  (D) to (E);           
    \draw[thick,<->]  (D) to (C);               
        
   \node () at (5,.3)  {\cite{ORS}} ;
  \node () at (2,-1)  {\cite{GORS}} ;  
  \node () at (8,-1)  {\cite{OY1}} ;
    \node () at (8,-2)  {\cite{GS2}} ;
   \node () at (-.6,-1.5)  {\cite{Ch}} ;
      \node () at (5,-3.3)  {\cite{SV2,Neg,GN}} ;
\end{tikzpicture}
\caption{The ordinary arrows represent theorems whereas the dotted arrows indicate conjectures. The relation between $\dim_{q,t}(e\overline L_{\frac{m}{n}})$ and $\chi_{\frakq,\frakt}(F_{m,n})$ represented by the dashed arrow is put on the rigorous footing only in the case of $m=kn+1$.}\label{fig:relation-torus}
\end{figure}

Overall relationships are summarized in Figure \ref{fig:relation-torus}. For the sake of brevity, we restrict ourselves to explain the relations at the level of $a=0$. To include general $a$-degrees requires technical details that are avoided in these notes. Thus, we refer the reader to original papers in Figure \ref{fig:relation-torus} for  general $a$-degrees. Let us mention that there is one arrow missing in Figure \ref{fig:relation-torus}, \textit{i.e.} the relation between the finite-dimensional module $\dim_{q,t}(e\overline L_{\frac{m}{n}})$ of the rational Cherednik algebra and the DAHA-superpolynomial $\overline\scP_{\yng(1)}^{\textrm{DAHA}}(T_{m,n};\fraka=0,\frakq,\frakt)$. Even though the rational Cherednik algebra  (rational DAHA) is a rational degeneration of DAHA as an algebra, the relation through torus knot homology has not been understood yet. We hope the reader will add one more arrow in Figure  \ref{fig:relation-torus} in near future. 

It is known that the HOMFLY homology is endowed with a rich algebraic structure. The authors of \cite{DGR} observed that the HOMFLY homology $\scH_{\yng(1)}(K)$ is equipped with the collection of anti-commuting differentials $\{d_N \}_{N \in \bZ}$  and the homology $H_*(\scH_{\yng(1)},d_N)$ with respect to $d_N$ is isomorphic to the $\fraksl(|N|)$ homology $\scH_{\fraksl(N),\yng(1)}(K)$. 
Later, the $d_N$ differentials have been constructed in terms of spectral sequences for $N>0$ \cite{Ras06}. Furthermore, colored HOMFLY homology $\scH_\lambda(K)$ is gifted with colored differentials $d_{\lambda\to\mu}$ \cite{GS12} so that the homology $H_*(\scH_{\lambda},d_{\lambda\to\mu})$ with respect to HOMFLY homology $\scH_{\mu}(K)$ colored by a lower color $|\mu|<|\lambda|$. It was noticed in \cite{GGS} that these structural properties become particularly manifest when one upgrades the triply-graded homology to a quadruply-graded homology. We review these results in \S\ref{sec:quadruple}. The study of the differential structure leads to better understanding of Habiro type structure for colored superpolynomials and we present new conjectures on cyclotomic expansions for
superpolynomials and HOMFLY polynomials colored by symmetric representation in this section. 

As we see above, there is a representation theoretic interpretation of HOMFLY homology of torus knot via the rational DAHA. In \cite{GOR,GORS}, the $d_N$ differentials are interpreted from the viewpoint of the rational DAHA.  Moreover, the approach of \cite{GORS} from the rational DAHA has been extended to colored cases in \cite{EGL}. Based on this construction, algebraic constructions of HOMFLY homology colored by symmetric representations have been proposed in \cite{GGS} and colored differentials have been elucidated by the rational DAHA. The section \S\ref{sec:matrix} discusses this approach and its relation to matrix factorizations.

Finally, let us stress that we do not intend to make these notes a thorough review of the most recent developments in knot homology and our perspective is very biased by our 
own research objectives. However, we will mention the most significant physical development in the last section \S\ref{sec:other}. The shortness of the last section by no means indicates the level importance
of the results but the level of competence of the survey in the related field.

\vspace{.5cm}

\subsection*{Convention}
Throughout this paper, we use the following skein relation for an unreduced HOMFLY polynomial
$\bar P_{\yng(1)}(K;a,q)$:
\bea\nonumber
&&a \bar P_{\yng(1)}\left({\raisebox{-.2cm}{\includegraphics[width=.6cm]{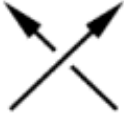}}}\right)
- a^{-1}\bar P_{\yng(1)}\left({\raisebox{-.2cm}{\includegraphics[width=.6cm]{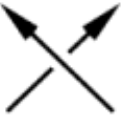}}}\right)
=
(q-q^{-1}) \bar P_{\yng(1)}\left({\raisebox{-.2cm}{\includegraphics[width=.6cm]{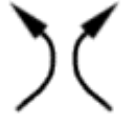}}}\right)\,,
\eea
with the unknot invariant
\be\nonumber
\bar P_{\yng(1)} ({\raisebox{-.1cm}{\includegraphics[width=.4cm]{unknot}}}) = {a - a^{-1} \over q -
q^{-1} }~. \ee
In addition, in this paper, a knot is always zero-framed and we do not consider non-trivial framings. 

\subsection*{Acknowledgement}
The authors are indebted to Johannes Walcher for organizing the workshop and the encouragement on writing this review. They are also grateful to Masaya Kameyama, Hitoshi Murakami and Paul Wedrich for carefully reading the manuscript and providing us valuable comments. S.N. would like to thank Sergei Gukov, P. Ramadevi, Ingmar Saberi, Marko St\v osi\'c, Piotr Sul{\l}kowski and Zodinmawia for the collaboration. The work of S.N. is supported by the ERC Starting Grant no.~335739 \textit{``Quantum fields and knot homologies''}, funded by the European Research Council under the European Union's Seventh Framework Programme. 
A.O. would like to thank Eugene Gorsky, Vivek Shende, Jake Rasmussen and Zhiwei Yun for the collaboration. The work of A.O. is partially supported by NSF CAREER grant DMS-1352398.

\section{Quadruply-graded homology}\label{sec:quadruple}
\subsection{Quadruply-graded homology}

In recent years, it has been proven that there exists rich structure tying together knot homology. Around the time that the definition of a triply-graded ($(a,q,t)$-graded) HOMFLY homology \cite{KR2} was given, it was predicted in \cite{DGR} that it is endowed  with a set of differentials $\{d_N\}_{N\in\bZ}$ where the homology with respect to $d_{N>0}$ is isomorphic to the $\fraksl(N)$ homology \cite{KR1}. Hence, the $d_N$ differentials open the passage from ``resolved conifold'' side to ``deformed conifold'' side.   Although both unreduced and reduced HOMFLY homology are endowed with the $d_N$ differentials, we only study properties of \emph{reduced} homology in this section:
\begin{itemize}\setlength{\parskip}{-0.1cm}
\item (Grading) The $(a,q,t)$-degree of the $d_N$ differential is \(\deg d_N=(-2,2N,-1)\).
\item (Involution) \(\scH(K)\) admits an involution 
\bea\label{involution}
\iota:( \scH_{\yng(1)}(K) )_{i,j,k} \xrightarrow{\cong}( \scH_{\yng(1)}(K)  )_{i,-j,k-j}~,
\eea
with the  property that \(\iota d_N = d_{-N} \iota\). 
\item (Anticommutativity) \(d_N d_M = -d_M d_N\) for \(N,M \in \bZ\). 
In particular, \(d^2_N = 0 \).
\item (Homology) The homology of \(d_{N>0}\) is isomorphic to reduced $\fraksl(N)$ homology:
\bea\label{dn}
H_*(\scH_{\yng(1)}(K),d_{N>0}) \cong \scH_{\fraksl(N),\yng(1)}(K)~.
\eea
 Moreover,  
\(H_*(\scH_{\yng(1)}(K), d_0)\) is isomorphic to the knot Floer homology 
\( \widehat{HFK}(K)\). 
\end{itemize}
In the sequel, it has been proven \cite{Ras06}   that for \(N>0\), there are spectral
sequences with \(E_1\) term \(\scH_{\yng(1)}(K)\) which converge to
\(\scH_{\fraksl(N),\yng(1)}(K)\). For \(N>0\), the conjecture is more or less equivalent to
the statement that these spectral sequences converge at the \(E_2\) term.

In \cite{GS12,GGS}, this approach has been extended to the colored case. Although the definition of colored HOMFLY homology is still very limited \cite{MSV} so far,  we conjecture the existence of the homology theory $\scH_{\lambda}(K)$ categorifying colored HOMFLY polynomials $P_{\lambda}(K)$. Then, the space of BPS states, refined Chern-Simons theory, representation theory of Lie superalgebra and the double affine Hecke algebra (DAHA) predict very rich structural  properties on $\scH_{\lambda}(K)$. We focus on rectangular Young diagrams $\lambda=(r^\rho)$, i.e. Young diagrams which have the form of a rectangle with $r$ rows and $\rho$ columns. This class includes both symmetric $(r)$ and anti-symmetric $(1^r)$ representations.

In particular, the structural properties become manifest when we introduce quadruple-gradings \cite{GGS}. Thus, let us explain the structural properties of colored HOMFLY homology of knots by using quadruple-gradings. The authors of \cite{GGS} managed to reconcile two different conventions for the homological ($t$-)grading so that colored HOMFLY homology turns into quadruply-graded $(\scH_{\lambda}(K))_{i,j,k,\ell}$: $(a,q,t_r,t_c)$-gradings: $t_r$ is the homological grading in the conventions of \cite{GS12}, whereas one can interpret $t_c$ as the one  in the conventions of \cite{AS,Ch,ITEP}. It is only in the uncolored case when the two homological gradings coincide and therefore the resulting homology is triply-graded in agreement with \cite{DGR}.

Moreover, it became apparent that all of the structural properties and isomorphisms
become particularly elegant once we replace the $q$-grading by the $Q$-grading defined by
\be\label{Q}
Q(x) := \frac{q(x)+t_r(x)-t_c(x)}{\rho} ~,
\ee
for every generator $x$ in $\scH_{(r^\rho)}(K)$. While it is just a regrading of $(\scH_{(r^\rho)}(K))_{i,j,k,\ell}$,  it is named the {\it tilde-version} of colored HOMFLY homology $\wt\scH_{(r^\rho)}(K)$ due to its importance:
\be\nonumber
(\wt\scH_{(r^\rho)}(K) )_{i,j,k,l} \; := \; (\scH_{(r^\rho)}(K))_{i,\rho j-k+l,k,l}~.
\ee
Since the two homological gradings are equal in the uncolored case, the $q$- and $Q$-grading of every generator of uncolored HOMFLY homology are the same.

Let also us introduce an important notion of knot homology. To every generator $x$ of the $(r)$-colored quadruply-graded HOMFLY homology,
one can associate a $\delta$-grading by
\bea\label{delta-grading-HOMFLY}
\delta(x):= a(x)+\frac{q(x)}{2}-\frac{t_r(x)+t_c(x)}{2}= a(x)+\frac{Q(x)}{2}-t_r(x)~.
\eea
Although the four gradings are independent in general, a knot $K$ is called \emph{homologically-thin} if all generators of $\scH_{(r)}(K)$ have the same $\delta$-grading which is equal to $\frac{r}{2}S(K)$ where we denote the $S$-invariant of the knot $K$ by $S(K)$  \cite{Ras04}. Otherwise, a knot is called \emph{homologically-thick} and the thick knot homology has more complicated structure \cite{DGR,GS12}. For instance, all two-bridge knots are homologically-thin. In contrast, the torus knots except $(2,2p+1)$ torus knots are homologically-thick \cite[Corollary 3]{Stosic}, and the examples of non-torus thick knots up to 10-crossings are as follows \cite{DGR}:
\be\nonumber
\bf 9_{42},\  10_{128},\ 10_{132},\ 10_{136},\ 10_{139},\ 10_{145},\ 10_{152},\ 10_{153},\ 10_{154},\ 10_{161}.
\ee

Although everything is conjectural, the definite advantage of the quadruply-graded theory is that it makes
all of the structural features and isomorphisms completely explicit.  To see them, we define the Poincar\'e polynomial of the quadruply-graded homology as
\bea\label{Poincare}
\scP_{(r^\rho)}(K;a,q, t_r,t_c)&:=& \sum_{i,j,k,\ell} a^i q^j t_r^k t_c^\ell ~\dim\;(\scH_{(r^\rho)}(K))_{i,j,k,\ell}~,\cr
\wt\scP_{(r^\rho)}(K;a,Q, t_r,t_c)&:=& \sum_{i,j,k,\ell} a^i Q^j t_r^k t_c^\ell ~\dim\;(\wt\scH_{(r^\rho)}(K))_{i,j,k,\ell}~,
\eea
where they are related by
\bea\label{Q-to-q-HOMFLY}
\wt\scP_{(r^\rho)}(K;a,q^\rho,t_rq^{-1},t_cq)=\scP_{(r^\rho)}(K;a,q,t_r,t_c)~.
\eea
Now, let us briefly describe the structural properties of the quadruply-graded colored HOMFLY homology. 

\begin{itemize}
\item {\bf Self-symmetry} \hfill \\
In the tilde-version of the colored HOMFLY homology, one can extend the involution \eqref{involution} to the colored cases: 
\bea\nonumber
(\wt\scH_{(r^\rho)}(K))_{i,j,k,\ell}&\cong&(\wt\scH_{(r^\rho)}(K))_{i,-j,k-\rho j,\ell-rj}~,
\eea
which can be stated at the level of the Poincar\'e polynomial 
\bea\label{self-symmetry} 
\wt\scP_{(r^\rho)}(K;a,Q,t_r,t_c)&=&\wt\scP_{(r^\rho)}(K;a,Q^{-1}t_r^{-\rho}t_c^{-r},t_r,t_c)~.
\eea
This becomes manifest only when we use the tilde-version of HOMFLY homology.

\item {\bf  Mirror/Transposition symmetry}  \hfill \\  
The $\lambda$-colored homology are related to the $\lambda^T$-colored homology by exchanging $t_r$ and $t_c$-gradings:
\bea\nonumber
(\wt\scH_{(\rho^r)}(K))_{i,j,k,\ell}\cong(\wt\scH_{(r^\rho)}(K))_{i,j,\ell,k}~,
\eea
which can be expressed in terms of the Poincar\'e polynomial 
\bea\label{mirror-HOMFLY}
\wt\scP_{(\rho^r)}(K;a,Q,t_r,t_c)&=&\wt\scP_{(r^\rho)}(K;a,Q,t_c,t_r)~.~~~~
\eea
In fact, the subscripts of two homological-gradings are the acronyms of "row" and "column".
This lifts the following relation between the colored HOMFLY polynomials
\bea
P_{\lambda^T}(K;a,q)=P_{\lambda}(K;a,q^{-1})~,
\label{PPmir}
\eea
for any representation $\lambda$ \cite{TVW,Zhu}.
\item {\bf Refined exponential growth property}  \hfill \\
Let $K$ be either a two-bridge knot or a torus knot. The $(r^\rho)$-colored quadruply-graded HOMFLY homology of the knot $K$ obeys the refined exponential growth property 
\bea
\wt\scP_{(r^\rho)}(K;a,Q,t_r,t_c=1)&=&\left[\wt\scP_{[1^\rho]}(K;a,Q,t_r,t_c=1)\right]^r\label{exp-growth-HOMFLY-1}~,\\
\wt\scP_{(r^\rho)}(K;a,Q,t_r=1,t_c)&=&\left[\wt\scP_{(r)}(K;a,Q,t_r=1,t_c)\right]^\rho\label{exp-growth-HOMFLY-2}~.
\eea
It follows immediately that
\bea\nonumber
\dim \scH_{(r^\rho)}(K)=\left[\dim \scH_{\yng(1)}(K)\right]^{r\rho}~.
\eea
It turns out that colored HOMFLY homology of the homologically-thick knot $\bf 9_{42}$ does not satisfy this property \cite[Appendix B]{GS12}. The analogous statement at the HOMFLY polynomial level \cite{Zhu} is that the following identity holds for any knot $K$ and an arbitrary representation $\lambda$
\bea\nonumber
P_{\lambda}(K;a,q=1)=\left[P_{\yng(1)}(K;a,q=1)\right]^{|\lambda|}~,
\eea
where $|\lambda|$ is the total number of the Young diagram corresponding to the representation $\lambda$.

\item  {\bf Colored differentials}  \hfill \\ 
For each rectangular Young diagram $(r^\rho)$, one can define colored differentials that remove any number of columns or rows from the original Young diagram $(r^\rho)$. For every $k$ with $r>k\ge 0$, there are two different column-removing  differentials  $d^\pm_{(r^\rho)\to(k^\rho)}$ with the  $(a, Q, t_r, t_c)$-degrees
 \bea\nonumber
\wt\deg ~d^+_{(r^\rho)\to (k^\rho)}&=&(-2,2,-1,-2k-1)\,, \cr
 \wt\deg ~d^-_{(r^\rho)\to (k^\rho)}&=&(-2,-2,-2\rho-1,-2r-2k-1)\,,
 \eea
 and for  every $\sigma$ with $\rho>\sigma\ge0$, there are two different row-removing differentials  $d^\pm_{(r^\rho)\to(r^\sigma)}$ on $\wt\scH_{(r^\rho)}(K)$ with the  $(a, Q, t_r, t_c)$-degrees,
\bea\nonumber
\wt\deg ~d^+_{(r^\rho)\to (r^\sigma)}&=&(-2,2,-2\sigma-1,-1)\,, \cr
 \wt\deg ~d^-_{(r^\rho)\to (r^\sigma)}&=&(-2,-2,-2\rho-2\sigma-1,-2r-1)\,.
\eea
The homology with respect to a colored differential is isomorphic to lower colored HOMFLY homology
\bea\label{colored-diff-sym-gen}
H_*(\wt\scH_{(r^\rho)}(K),d^\pm_{({r^\rho})\to ({k^\rho})})&\cong& \wt\scH_{(k^\rho)}(K)~,\cr
H_*(\wt\scH_{(r^\rho)}(K),d^\pm_{({r^\rho})\to ({r^\sigma})})&\cong& \wt\scH_{(r^\sigma)}(K)~.
\eea
Precisely speaking, the isomorphisms above involve regrading. Although the quadruple-gradings makes the regrading very explicit, the details are not used in this paper, so we refer to \cite[\S3]{GGS}.

\item  {\bf Universal colored differentials}  \hfill \\ 
When the color is specified either by a Young diagram $(r,r)$ or by $(2^r)$,  there exists yet another set of colored differentials $d^\uparrow$ or $d^\leftarrow$ with  $(a, Q, t_r, t_c)$-degrees
\bea\nonumber
\wt\deg ~d^\uparrow=(0,0,-2,0)~, \quad \wt\deg ~d^\leftarrow=(0,0,0,2)~,
\eea
 so that 
\bea\nonumber
H_*(\wt\scH_{(r,r)}(K),d^\uparrow)&\cong& \wt\scH_{(r)}(K)~,\cr
H_*(\wt\scH_{(2^r)}(K),d^\leftarrow)&\cong& \wt\scH_{(1^r)}(K)~.
\eea
They are called universal colored differentials because they are universal in the sense that their $a$-degree is equal to $0$. We also refer to \cite[\S3]{GGS} for the regrading used in these isomorphisms. 

\item  {\bf $\fraksl(m|n)$ differentials}  \hfill \\ 
It has been revealed that the existence of the $d_N$ differential and the colored differentials comes from the representation theory of the Lie superalgebras $\fraksl(m|n)$. In fact, these differentials can be expressed as $\{d_{m|n}\}$ labeled by two non-negative integers $(m,n)$ associated to the Lie superalgebra $\fraksl(m|n)$. For instance, the colored differentials can be written as
\bea\nonumber
d^+_{(r^\rho)\to(k^\rho)} =d_{\rho|k}~,&\qquad& d^+_{(r^\rho)\to(r^\sigma)} =d_{\rho+\sigma|0}~,\cr
d^-_{(r^\rho)\to(k^\rho)} =d_{0|r+k}~,&\qquad& d^-_{(r^\rho)\to(r^\sigma)} =d_{\sigma|r}~.
\eea
Representations of Lie superalgebras are also labelled by Young diagrams as in the case of representations of ordinary Lie algebras. For $r>k\ge0$, isomorphisms of the $\fraksl(\rho|k)$-representations give rise to the positive column-removing differentials whereas the isomorphisms of the $\fraksl(0|r+k)$-representations give rise to the negative column-removing differentials:
\bea\nonumber
&&(r^\rho)\cong(k^\rho) \qquad \textrm{as} \ \fraksl(\rho|k) \ \textrm{representations}\cr
&&(r^\rho)\cong(k^\rho) \qquad \textrm{as} \ \fraksl(0|r+k) \ \textrm{representations}~.
\eea
The row-removing differentials can be also related to representations of Lie superalgebras. In fact, for two Young diagrams $\lambda$ and $\mu$, we have the mirror/transposition symmetry in the representations of Lie superalgebras:
\bea\label{supergroup-mirror}
\lambda\cong\mu &\quad&\textrm{as} \ \fraksl(m|n) \ \textrm{representations}\cr
\textrm{if and only if}\qquad\lambda^T\cong\mu^T &\quad&\textrm{as} \ \fraksl(n|m) \ \textrm{representations}~.
\eea
In the case of the fundamental representations, the $d_N$ differentials can be written as $d_N=d_{N|0}$ for $N>0$, $d_N=d_{0|N}$ for $N<0$, and $d_0=d_{1|1}$. Therefore, the differentials $d_{m|n}$ are generalizations of the differentials $\{d_N\}$ \eqref{dn}. 

It is worth mentioning that the properties of quantum invariants for $U_q(\fraksl(m|n))$ have been investigated in \cite{TVW}. It was shown in \cite{TVW} that $U_q(\fraksl(m|n))$ colored quantum invariants stabilize to colored HOMFLY polynomials and the mirror/transposition symmetry \eqref{PPmir} follows from the symmetry \eqref{supergroup-mirror}.

\end{itemize}

As we will see in \S\ref{sec:DAHA}, the mirror/transposition symmetry \eqref{mirror-HOMFLY} and the refined exponential growth property (\ref{exp-growth-HOMFLY-1},\ref{exp-growth-HOMFLY-2}) have been proven for the DAHA-superpolynomials of torus knots in \cite{Ch2}. In addition, we will see in \S\ref{sec:matrix} that the colored differentials are conjecturally constructed from the viewpoint of the rational DAHA \cite{GGS}. However, we should emphasize that all these properties are still at the level of conjectures. Above all, the definition of quadruply-graded homology as well as physical interpretation of quadruply-gradings are expected to be given.

The colored Kauffman homology is gifted with as rich a structure as colored HOMFLY homology. It also admits quadruple-gradings as well as colored differentials, $\frakso/\fraksp$ differentials, and universal (colored) differentials. More interestingly, the colored Kauffman homology includes colored HOMFLY homology. For more detail, we refer the reader to \cite{GW,Nawata:2013mzx}. In addition, recently, the differential structure on the knot homology with exceptional groups has been investigated from the viewpoint of DAHA \cite{Elliot:2015pra}. When the color is specified by non-rectangular Young tableaux, some of the properties do not hold. It is an important open problem to understand the properties of HOMFLY homology colored by non-rectangular Young diagrams.

\subsection{Cyclotomic expansions}
Habiro has found an interesting expansion of colored Jones polynomials, known as the cyclotomic expansion.
\begin{theorem}\text{\cite{Habiro:2008}}\label{Habiro-thm}
For a knot $K$, there exists a function
\be
c(K,N=2):\bZ_{\ge0}\longrightarrow\bZ[q^\pm] \ ; \ k\mapsto c_k(K,N=2;q)~,\nonumber
\ee
such that reduced colored Jones polynomial of the knot $K$ is expressed as
\be\nonumber
J_{\fraksl(2),(r)}(K;q)=\sum_{k\ge0}c_k(K,N=2;q)(q^{2r+4};q^2)_k(q^{-2r};q^2)_k~.
\ee
\end{theorem}
In fact, the structural properties of colored HOMFLY homology strongly dictates the form of its Poincar\'e polynomial, which remarkably provides a new perspective to cyclotomic expansions of Habiro type for knot invariants. Thus, in this subsection, we shall present a conjecture for a cyclotomic expansions of superpolynomials colored by symmetric representations and its implications for colored HOMFLY polynomials and $\fraksl(N)$ quantum invariants.

As explained in the previous subsection, all the generators except one in uncolored HOMFLY homology can be paired by the canceling differential $d^-_{\yng(1)\to 0}$ whose $(a,q,t)$-degree is $(-2,-2,-3)$, which can be seen as at the level of the Poincar\'e polynomial
\bea\nonumber
\scP_{\yng(1)}(K;a,q,t)=A_1 + (A_2+\cdots +A_m)(1+a^2q^{2}t^3)~,
\eea
where $A_i=a^{\ell}q^{m}t^n$ for $\ell,m,n\in \bZ$ and $A_1$ represents the one remaining generator. Then, the $t_c=1$ specialization of Poincar\'e polynomial of $(r)$-colored homology can be determined by the refined exponential growth property \eqref{exp-growth-HOMFLY-1}
\bea
\wt\scP_{(r)}(K;a,Q,t_r,t_c=1)&=&(\scP_{\yng(1)}(K;a,Q,t_r))^r\cr
&=&\sum_{r\ge k\ge 0} A_1^{r-k}(1+a^2Q^{2}t_r^3)^k   \binom{r}{k} \sum_{i_2+\cdots+i_m=k} A_2^{i_2} \cdots A_m^{i_m}   \binom {k} {i_2,\ldots,i_m}~.\nonumber
\eea
To get  the complete Poincar\'e polynomial, the $(1+a^2Q^{2}t_r^3)^k$ is replaced by $t_c^2$-Pochhammer symbol $(-a^2Q^{2}t_r^3t_c^{2r+1};t_c^2)_k$ whereas we substitute the $t_c^2$-binomial $t_c^{-2rk}{r \brack k}_{t_c^2}$ for  the binomial $\binom{r}{k}$. Remarkably, it turns out that the second summation is independent of the color $r$  in all the examples we know even after restoring $t_c$-gradings. Therefore, we conjecture that there exists a cyclotomic expansion of superpolynomial colored by a rank-$r$ symmetric representation $(r)$ for a two-bridge knot or torus knot $K$.
\begin{conjecture}\label{conj-super}
Let $K$ be either a two-bridge knot or a torus knot. Then, there exists a function
 \be
\scC(K):\bZ_{\ge0}\longrightarrow\bZ[a^\pm,Q^\pm,t_r^\pm,t_c^\pm] \ ; \ k\mapsto \scC_k(K;a,Q,t_r,t_c)~,\nonumber
\ee
with $\scC_{k=0}(K;a,Q,t_r,t_c)=1$ such that Poincar\'e polynomial of $(r)$-colored quadruply-graded HOMFLY homology of the knot $K$ is written as 
\be\label{cyclotomic-super}
\wt\scP_{(r)}(K;a,Q,t_r,t_c)=( \pm a^{\bullet}Q^{\bullet}t_r^{\bullet}t_c^{\bullet})^r \sum_{r\ge k\ge0}\scC_k(K;a,Q,t_r,t_c) t_c^{-2rk}\left(-a^2 Q^{2} t_r^{3} t_c^{2r+1};t_c^2\right)_k{r \brack k}_{t_c^2}
\ee
where $a^{\bullet}Q^{\bullet}t_r^{\bullet}t_c^{\bullet}$ means an appropriate shift of $(a,Q,t_r,t_c)$-degree. \end{conjecture}
Indeed, the Poincar\'e polynomials of the $(2,2p+1)$, $(3,4)$ torus knot and the (double) twist knots  take this form  \cite{Fuji:2012pm,Fuji:2012nx,Fuji:2012pi,Nawata:2012pg,GNSS}. As we will see below, the colored superpolynomials of the knots $\bf 6_2$ and $\bf 6_3$  also allow this cyclotomic expansion.

Since this structure will be reflected at the level of HOMFLY polynomials, we also conjecture the cyclotomic expansions of HOMFLY polynomials colored by symmetric representations for \emph{any} knots.
\begin{conjecture}\label{HOMFLY-cyclotomic}
For a knot $K$, there exists a function
 \be
C(K):\bZ_{\ge0}\longrightarrow\bZ[a^\pm,q^\pm] \ ; \ k\mapsto C_k(K;a,q)~,\nonumber
\ee
which satisfies the following properties
\begin{itemize}\setlength{\parskip}{-0.1cm}
\item $C_{k=0}(K;a,q)=1$
\item for positive integers $N>0$ and $k>0$, $C_{k}(K;a=q^N,q)$ is divisible by $(q^2;q^2)_k$
\be\label{divisivility}\frac{C_{k}(K;a=q^N,q)}{(q^2;q^2)_k} \in \bZ[q^\pm]~,\ee
\end{itemize}
such that the $(r)$-colored reduced HOMFLY polynomial of the knot $K$ is expressed as
\bea\label{cyclotomic-HOMFLY}
P_{(r)}(K;a,q)
&=&(a^{\bullet}q^{\bullet})^r\sum_{r\ge k\ge0} C_k(K;a,q) q^{-2rk}\left(a^2 q^{2r};q^2\right)_k{r \brack k}_{q^2}~.
\eea
\end{conjecture}
 In fact, a similar conjecture for HOMFLY polynomials was first stated in \cite{Kononov:2015dda}, and Conjecture \ref{HOMFLY-cyclotomic} is a more precise statement. Note that the formula \eqref{cyclotomic-HOMFLY} is obtained from the de-categorifying substitution in \eqref{cyclotomic-super}  $$\wt\scP_{(r)}(K;a,Q=q,t_r=-q^{-1},t_c=q)=P_{(r)}(K;a,q)~,$$  even though Conjecture \ref{HOMFLY-cyclotomic} is applied for any knot $K$.  The second property on $C(K)$ means that although $C_k(K;a,q) /(q^2;q^2)_k$ is a rational function of $a,q$ in general for a positive integer $k$, it becomes a Laurent polynomial of $q$ once one substitutes $a=q^N$ for any positive integer $N$.
  Importantly, this property \eqref{divisivility} relates Conjecture \ref{HOMFLY-cyclotomic} to Theorem \ref{Habiro-thm} of Habiro. Since we can write the factor in the summand of \eqref{cyclotomic-HOMFLY} as
$$q^{-2rk}{r \brack k}_{q^2}=(-1)^k q^{(1-k)k}\frac{(q^{-2r};q^2)_k}{(q^2;q^2)_k}~,$$
the property \eqref{divisivility} tells that the denominator $(q^2;q^2)_k$ can be absorbed into $C(K)$ for $\fraksl(N)$ quantum invariants colored by symmetric representations. Hence, substituting $a=q^N$, the immediate consequence of Conjecture \ref{HOMFLY-cyclotomic} is as follows:

\begin{conjecture}\label{}
For a knot $K$, there exists a function
 \be
c(K,N):\bZ_{\ge0}\longrightarrow\bZ[q^\pm]  \ ; \ k\mapsto c_k(K,N;q)~,\nonumber
\ee
with $c_{k=0}(K,N;q)=1$ such that the $(r)$-colored $\fraksl(N)$ quantum invariant of the knot $K$ is expressed as
\bea\nonumber
J_{\fraksl(N),(r)}(K;q)
&=&(q^{\bullet})^r\sum_{r\ge k\ge0} c_k(K,N;q) \left( q^{2N+2r};q^2\right)_k\left(q^{-2r};q^2\right)_k~.
\eea
\end{conjecture}
At $N=2$, this is nothing but Theorem \ref{Habiro-thm} of Habiro.  The same conjecture is independently stated in \cite[Conjecture 1.3]{CLZ} for $\fraksl(N)$ quantum invariants.

\begin{figure}[H]
\centering
\includegraphics[width=\textwidth]{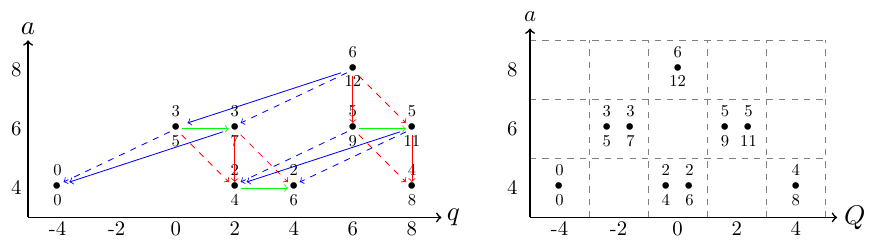}
\caption{The homology diagram for the reduced $(2)$-colored HOMFLY homology of the trefoil. A dot represents a generator in the homology and its $a$- and $q$-gradings are scaled in the vertical and horizontal lines, respectively. The $t_r$- and  $t_c$-gradings are written above and below the dot, respectively. Here, the canceling differentials $d^+_{(2)\to(0)}=d_{1|0}$ and $d^-_{(2)\to(0)}=d_{0|2}$ are drawn by the red and blue dashed arrow. In addition, the red and blue arrows represent the colored differential $d^+_{(2)\to(1)}=d_{1|1}$ and $d^-_{(2)\to(1)}=d_{0|3}$, respectively, while the green arrow shows the action of the universal colored differential. The self-symmetry which is the symmetry along the vertical axis at $Q=0$ becomes manifest in the right diagram.}
\label{fig:quad-trefoil}
\end{figure}

\subsection{Examples}

Let us look at the example of the trefoil $T_{2,3}$. The Poincar\'e polynomial of the uncolored HOMFLY homology can be written as
\be\label{31-1}
\scP_{\yng(1)}(T_{2,3};a,q,t)=a^2q^{-2}+a^2q^2t^2+a^4t^3.
\ee
It is easy to see that the $\delta$-gradings of all the three generators are $+1$ so that it is homologically-thin. Beside, the $(2)$-colored superpolynomial can be written as
\bea\label{quad31}
\scP_{\yng(2)}(T_{2,3};a,q,t_r,t_c)&=& a^4(q^{-4} +q^2t_r^2t_c^4+q^4t_r^2t_c^6 +q^8t_r^4t_c^8) \cr
&&+  a^6 (t_r^3t_c^5+q^2t_r^3t_c^7 + q^6t_r^5t_c^9+q^8t_r^5t_c^{11}) +a^8q^6t_r^6t_c^{12}.
\eea
where we combine the result with $t_r$-grading in \cite{GS12} and the one with $t_c$-grading \cite{AS,ITEP}. In the left of Figure \ref{fig:quad-trefoil}, the colored differentials in the homology are depicted. By a simple change of variables \eqref{Q}, the tilde version of HOMFLY homology follows from \eqref{Q} and its Poincar\'e polynomial is 
\bea\label{tilquad31}
\wt{\scP}_{\yng(2)}(T_{2,3};a,Q,t_r,t_c)&=& a^4(Q^{-4} +t_r^2t_c^4+t_r^2t_c^6 +Q^4t_r^4t_c^8) \\
&&+a^6 (Q^{-2}t_r^3t_c^5+Q^{-2}t_r^3t_c^7 + Q^2t_r^5t_c^9+Q^2t_r^5t_c^{11}) +a^8t_r^6t_c^{12}.\nonumber
\eea
It is easy to check the refined exponential growth property
\be\nonumber
\wt{\scP}_{\yng(2)}(T_{2,3};a,Q,t_r,t_c=1)=\left[\scP_{\yng(1)}(T_{2,3};a,Q,t_r)\right]^2~,
\ee
as well as the self-symmetry
\[
\wt{\scP}_{\yng(2)}(T_{2,3};a,Q^{-1}t_r^{-1}t_c^{-2},t_r,t_c)=\wt{\scP}_{\yng(2)}(T_{2,3};a,Q,t_r,t_c)~.
\]
For more detail, we refer to \cite[\S3]{GGS}.

Using the structural properties, cyclotomic expansions of superpolynomials colored by symmetric representations have been obtained for $(2,2p+1)$, $(3,4)$ torus knots, twist knots, and double twist knots so far \cite{Fuji:2012pm,Fuji:2012nx,Fuji:2012pi,Nawata:2012pg,GNSS}. Here we present new examples: colored superpolynomials of the $\bf 6_2$ and $\bf 6_3$ knots. The uncolored superpolynomials of these knots can be written \cite{DGR} as 
\bea\nonumber
\scP_{\yng(1)}({\bf 6_2};a,q,t)&=&-t^{-1}+(q^{-2}t^{-2}+t^{-1}+q^2)(1+a^2q^{-2}t)(1+a^2q^2t^3)~,\cr
\scP_{\yng(1)}({\bf 6_3};a,q,t)&=&1+a^{-2}(q^{-2}t^{-3}+t^{-2}+q^2t^{-1})(1+a^2q^{-2}t)(1+a^2q^2t^3)~.
\eea
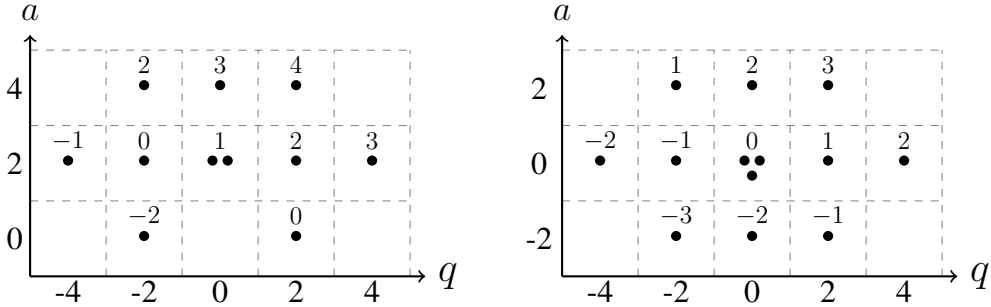
\begin{figure}[h]\centering
\begin{tikzpicture}
\draw[step=1cm,gray,dashed,very thin] (-5,-1) grid (0,2);
  \draw[thick,->] (-5,-1) -- (.2,-1);
\draw[thick,->] (-5,-1) -- (-5,2.2);
\node at (-4.5,-1.2) {-4};
\node at (-3.5,-1.2) {-2};
\node at (-2.5,-1.2) {0};
\node at (-1.5,-1.2) {2};
\node at (-.5,-1.2) {4};
\node at (.5,-1) [scale =1.3]{$q$};
\node at (-5.2,-.5) {0};
\node at (-5.2,0.5) {2};
\node at (-5.2,1.5) {4};
\node at (-5,2.5) [scale =1.1] {$a$};
\node at (-4.5,.5) [scale =3]{$\cdot$};
\node  at (-4.5,0.8) [scale =.8]{$-1$};
\node at (-3.5,0.5) [scale =3]{$\cdot$};
\node  at (-3.5,0.8) [scale =.8]{$0$};
\node at (-2.4,0.5) [scale =3]{$\cdot$};
\node at (-2.6,0.5) [scale =3]{$\cdot$};
\node  at (-2.5,0.8) [scale =.8]{$1$};
\node at (-1.5,0.5) [scale =3]{$\cdot$};
\node  at (-1.5,0.8) [scale =.8]{$2$};
\node at (-.5,0.5) [scale =3]{$\cdot$};
\node  at (-.5,0.8) [scale =.8]{$3$};
\node at (-3.5,1.5) [scale =3]{$\cdot$};
\node  at (-3.5,1.8) [scale =.8]{$2$};
\node at (-2.5,1.5) [scale =3]{$\cdot$};
\node  at (-2.5,1.8) [scale =.8]{$3$};
\node at (-1.5,1.5) [scale =3]{$\cdot$};
\node  at (-1.5,1.8) [scale =.8]{$4$};
\node at (-3.5,-.5) [scale =3]{$\cdot$};
\node  at (-3.5,-.2) [scale =.8]{$-2$};
\node at (-1.5,-.5) [scale =3]{$\cdot$};
\node  at (-1.5,-.2) [scale =.8]{$0$};

\draw[step=1cm,gray,dashed,very thin] (2,-1) grid (7,2);
  \draw[thick,->] (2,-1) -- (7.2,-1);
\draw[thick,->] (2,-1) -- (2,2.2);
\node at (2.5,-1.2) {-4};
\node at (3.5,-1.2) {-2};
\node at (4.5,-1.2) {0};
\node at (5.5,-1.2) {2};
\node at (6.5,-1.2) {4};
\node at (7.5,-1) [scale =1.3]{$q$};
\node at (1.7,-.5) {-2};
\node at (1.7,0.5) {0};
\node at (1.7,1.5) {2};
\node at (2,2.5) [scale =1.1] {$a$};
\node at (2.5,.5) [scale =3]{$\cdot$};
\node  at (2.5,0.8) [scale =.8]{$-2$};
\node at (3.5,0.5) [scale =3]{$\cdot$};
\node  at (3.5,0.8) [scale =.8]{$-1$};
\node at (4.4,0.5) [scale =3]{$\cdot$};
\node at (4.6,0.5) [scale =3]{$\cdot$};
\node at (4.5,0.3) [scale =3]{$\cdot$};
\node  at (4.5,0.8) [scale =.8]{$0$};
\node at (5.5,0.5) [scale =3]{$\cdot$};
\node  at (5.5,0.8) [scale =.8]{$1$};
\node at (6.5,0.5) [scale =3]{$\cdot$};
\node  at (6.5,0.8) [scale =.8]{$2$};
\node at (3.5,1.5) [scale =3]{$\cdot$};
\node  at (3.5,1.8) [scale =.8]{$1$};
\node at (4.5,1.5) [scale =3]{$\cdot$};
\node  at (4.5,1.8) [scale =.8]{$2$};
\node at (5.5,1.5) [scale =3]{$\cdot$};
\node  at (5.5,1.8) [scale =.8]{$3$};
\node at (3.5,-.5) [scale =3]{$\cdot$};
\node  at (3.5,-.2) [scale =.8]{$-3$};
\node at (4.5,-.5) [scale =3]{$\cdot$};
\node  at (4.5,-.2) [scale =.8]{$-2$};
\node at (5.5,-.5) [scale =3]{$\cdot$};
\node  at (5.5,-.2) [scale =.8]{$-1$};
\end{tikzpicture}
\caption{uncolored HOMFLY homology of $\bf 6_2$ (left) and $\bf  6_3$ (right) knot} \label{fig:62-63}
\end{figure}
As Figure \ref{fig:62-63} shows, the homologies of the  the $\bf 6_2$ and $\bf 6_3$ knots are very similar. Then, using the multinomial formula, the refined exponential growth property \eqref{exp-growth-HOMFLY-1} tells us that the $t_c=1$ specialization of the Poincar\'e polynomials of $(r)$-colored HOMFLY homology are given by
\bea\nonumber
\wt\scP_{(r)}({\bf 6_2};a,Q,t_r,t_c=1)&=&\sum _{r\ge k\ge j \ge i \ge0} (-t_r^{-1})^{r-k} (Q^{-2}t_r^{-2})^{k-j} t_r^{-(j-i)}Q^{2i}
   \binom{r}{k}  \binom{k}{j}  \binom{j}{i}\cr
   &&\hspace{3cm} (1+a^2 Q^{-2}t_r)^k(1+a^2 Q^2 t_r^3)^k\cr
\wt\scP_{(r)}({\bf 6_3};a,Q,t_r,t_c=1)&=&\sum _{r\ge k\ge j \ge i \ge0} a^{k} (Q^{-2}t_r^{-3})^{j - i} t_r^{-2i}(Q^2t_r^{-1})^{2(k - j)}
   \binom{r}{k}  \binom{k}{j}  \binom{j}{i} \cr
   &&\hspace{3cm}(1+a^2 Q^{-2}t_r)^k(1+a^2 Q^2 t_r^3)^k
\eea
The $t_c$-degrees have to be determined in such a way that the homology is endowed with all the differentials. Nevertheless, roughly speaking, the binomials are replaced by $t_c^2$-binomials whereas a product $(1+x)^\ell$ is replaced by a $t_c^2$-Pochhammer symbol $(-xt_c^*;t_c^2)_\ell$. In these examples, closed formulas can be written as
\bea
\wt\scP_{(r)}({\bf 6_2};a,Q,t_r,t_c)&=&\sum _{r\ge k\ge j \ge i \ge0} (-1)^{r-k} Q^{2 (i+j-k)}
t_r^{i+j-k-r} t_c^{i^2-j^2+2 j k-2 k r+k-r}
   {r \brack k}_{t_c^2}  {k \brack j}_{t_c^2} {j \brack i}_{t_c^2} \cr
   &&\left(-a^2 Q^{-2}t_r t_c;t_c^2\right)_j \left(-a^2 Q^{-2}t_rt_c^{2 i+1};t_c^2\right)_{k-j} \left(-a^2 Q^2 t_r^3t_c^{2 r+1};t_c^2\right)_k ~,\cr
\wt\scP_{(r)}({\bf 6_3};a,Q,t_r,t_c)&=&\sum _{r\ge k\ge j \ge i \ge0} a^{-2 k} Q^{2 (i - 2 j + k)}
   t_r^{i - 2 j - k}t_c^{i^2 + k (-2 j + k - 2 r)}{r \brack k}_{t_c^2}  {k \brack j}_{t_c^2} {j \brack i}_{t_c^2} \cr
   &&\left(-a^2 Q^{-2}t_r t_c;t_c^2\right)_j \left(-a^2 Q^{-2}t_rt_c^{2 i+1};t_c^2\right)_{k-j} \left(-a^2 Q^2 t_r^3t_c^{2 r+1};t_c^2\right)_k ~.\nonumber
\eea
It is easy to see that the colored superpolynomials of the knot ${\bf 6_2}$ and ${\bf 6_3}$  take the form of the cyclotomic expansion in Conjecture \ref{conj-super}.

\section{Knot homology and planar curves}\label{sec:planar}
In this section, we explain the relationships of HOMFLY homology, nested Hilbert schemes of a planar curve and the moduli space of stable pairs. In particular. we shall state the original conjectures from \cite{OS, ORS} and explain the version of the conjecture for the colored knot invariants \cite{DHS}. These conjectures provide rigorous formulations of the geometric transition.

\subsection{Knot homology and Hilbert schemes of planar curve}\label{subsec:geometry}

Let $C=\{E(x,y)=0\}\subset \CC^2$ be a planar curve.  Then $C^{[n]}$ stands for the Hilbert scheme of $n$ points
on $C$,  that is, the set of ideals $I\subset \CC[x,y]$ that contain $E$ and have codimension $n$. If $C$ is smooth, 
the Hilbert scheme is the $n$-th symmetric power of the curve; for a singular curve it is a partial resolution
of the symmetric power.
If we assume that $E(0,0)=0$, then  $C^{[n]}_{(0,0)}$ is the punctual Hilbert scheme (i.e. the moduli space of ideals defining 
a fat point supported at $(0,0)$). Algebraically, it is the set of ideals from $C^{[n]}$ that contains $x^N,y^N$ for some $N$.
Let us introduce the following  nested Hilbert scheme:
$$C_{(0,0)}^{[l]} \times C_{(0,0)}^{[l+m]} \supset 
C_{(0,0)}^{[l,\,l + m]} := \{(I,J) |  
I \supset J \supset I\cdot (x,y)\}$$
At the $m=0$ specialization, it reduces to the Hilbert scheme. In general, $C_{(0,0)}^{[l,l+m]}$ maps to $C_{(0,0)}^{[l]}$  with smooth fibers that 
are constant over the locus of ideals with fixed minimal number of generators. It is therefore possible to restate our conjecture in
terms of Euler characteristics of loci with a fixed minimal number of generators \cite{OS}.
 Below we state a conjecture relating the topological invariants of the nested Hilbert schemes of a planar curve to HOMFLY invariants of 
 the  link for  the singularity of the   curve.

The links $L_{C,{(0,0)}}$ that constitute the intersection of the curve $C$ with the small $3$-sphere around $(0,0)$ are called {\it algebraic}. When $E=x^m-y^n$,  the link
is the torus link $T_{m,n}$. The following formula was conjectured in \cite{OS} and proved for the torus knot cases and some $2$-cablings of the trefoil; it is also shown in \cite{OS} that 
the case $a=-1$ of the conjecture is equivalent to the previously known theorem of \cite{GZCD};
 the proof for arbitrary 
plane curve singularity is given by Maulik \cite{Mau}.

\begin{theorem} \text{\cite{Mau,OS,GZCD}}
  \label{conj:homfly} Let
   $\mu=\dim \C[[x,y]]/(\frac{\partial E}{\partial x},\frac{\partial E}{\partial y})$ be the Milnor number
  of the singularity at $(0,0)$.  Then, we have
  \begin{equation*}
    \bar{P}_{\yng(1)}(L_{C,(0,0)})  =   (a/q)^{\mu-1} \sum_{l,m}
    q^{2l} (-a^2)^m \chi(C_{(0,0)}^{[l,\,l + m]})~.
  \end{equation*}
\end{theorem}

As we mentioned above the proof of the most general case of the formula could be found in \cite{Mau}. The proof proceeds by considering more general conjecture that involves the colored knot invariants. In the next subsection, we shall provide more details
on the colored conjecture.

To state its homological version, let us denote the Poincar\'e polynomial of 
the triply graded  HOMFLY homology $(\overline{\mathscr{H}}_{\yng(1)}(K))_{i.j.k}$
defined by Khovanov and Rozansky \cite{KR2}. 
$$\overline\scP_{\yng(1)}(K;a,q,t) := \sum_{i,j,k} a^i q^j t^k ~\dim(\overline{\mathscr{H}}_{\yng(1)}(K))_{i,j,k}~,$$
like \eqref{Poincare}. The $t=-1$ specialization is the $(a,q)$-graded Euler characteristics of HOMFLY homology that is  indeed 
HOMFLY invariant   $\overline{P}_{\yng(1)} (K;a,q)= \overline \scP_{\yng(1)}(K;a,q,t=-1)$. Then, using the same convention above,
the statement of the homological version is given in Conjecture \ref{planar-homology}.

When the curve admits a $\CC^*$-action, a combinatorial formula for the algebro-geometric side
of Conjecture \ref{planar-homology} has been derived in \cite{ORS}. The combinatorics of the Hilbert scheme is much easier than the combinatorics 
of the homological algebra underlying the definition of HOMFLY homology \cite{KR2}. In particular, the computation of the HOMFLY homology of torus knots directly from the definition in \cite{KR2} appears to be tedious \cite{CM} and no rigorous calculations have been given for the torus knots $T_{m,n}$ with $n>3$ whereas the combinatorics of the Hilbert scheme yields explicit formulas of the RHS in Conjecture \ref{planar-homology} for torus knots. Therefore, Conjecture \ref{planar-homology} has been checked for torus knots $T_{2,n}$, $T_{3,n}$. As we will see in \S \ref{sec:rational-DAHA}, the relation between HOMFLY homology and the Hilbert scheme of a planar curve reveals the connections of
 HOMFLY homology to representation theory of the rational DAHA for torus knots and leads to rigorous formulation of the geometric engineering.

\subsection{Colored invariants and stable pairs}\label{subsec:colored pairs}

When one searches for a moduli space  that would match with the colored HOMFLY invariants $\bar{P}_{\lambda}$, the naive guess might be the moduli space of the ideals on the thick curve   in $\CC^3=\{(x,y,z)\}$ 
$$C_{\lambda}=\{ E^{\lambda_1}(x,y)=0, zE^{\lambda_2}(x,y)=0,\dots\}$$
that has a fat point of shape $\lambda$ as a generic cross-section. It turns out that this moduli space is not suitable, but the following close cousin passes numerical tests.
The moduli space $\mathcal{P}^\lambda_n(rC)_{(0,0)}$ consists of pairs of a pure sheaf $\mathcal{G}$ with support on $C$ 
and a map $s$ surjective outside $(0,0)$ such that: 
$$ [\mathcal{O}_{\CC^3}\xrightarrow{s}\mathcal{G}]\in \mathcal{P}^\lambda_n(rC)_{(0,0)}\quad \mbox{ iff } \quad \textrm{Ker}(s)=(E^{\lambda_1},zE^{\lambda_2},\dots)~.$$

This moduli space appears naturally in  the study of moduli spaces of pairs \cite{PT1}. When 
one counts curves in a Calabi-Yau threefold that are  homologous to $\beta\in H_2(X)$,
it is generally expected (and shown in some cases \cite{PT2}) that  the count 
is given in terms of so-called BPS states, which mathematically manifest themselves as topological
invariants of the moduli spaces of sheaves on the singular curves  that are homologous to $\beta$.

In the case of $\lambda=(m)$, we deal with sheaves on the fat but still planar 
curve $mC$. Then,  \cite[Appendix B]{PT2} contains the proof of 
$ \mathcal{P}_n^{(m)}(C)_{(0,0)}=(C^{[n]}_{(m)})_{(0,0)}$
where $C_{(m)}$ is a planar curve $E^m=0$,  \textit{i.e.} the $m$-fattening of $C$.
On the other hand, for $\lambda=(1^m)$, one can immediately identify the moduli space with the $m$-step nested  Hilbert scheme. An general partition is a hybrid of these cases.

The first observation between colored knot invariants and the moduli space of pairs has been made in \cite{O-ober} involving colored $\fraksl(\infty)$ quantum invariants. From the cabling formula \cite{Cabling}, there are unique powers $f(\lambda,K)$, $g(\lambda,K)$ for an algebraic knot $K$  such that 
$ q^{f(\lambda,K)}a^{g(\lambda,K)} \bar{P}_\lambda(K;a=0,q=0)=1$.
Let us define the $\fraksl(\infty)$ quantum invariant  by 
$$\bar{P}_{\fraksl(\infty),\lambda}(K;q):=q^{f(\lambda,K)}a^{g(\lambda,K)}\bar{P}_\lambda(K;{a=0},q)~.$$
Then, it was conjectured in \cite{O-ober} that  if $L_{C,(0,0)}$ be a link  of singularity of $C$ at $(0,0)$, then
\be\label{O-ober-eq}
 \bar{P}_{\fraksl(\infty),\lambda}(L_{C,(0,0)})=\sum_{n} \chi(\mathcal{P}^{\lambda}_n(C)_{(0,0)}) q^{2n}.
 \ee

An extension of the above formula for colored HOMFLY invariant was investigated in \cite{DHS}.
Let $X=\mathcal{O}(-1)\oplus\mathcal{O}(-1)\to \mathbb{P}^1$ be the resolved conifold and $\mathbb{C}^2\simeq D\subset X$ is the fiber over $0\in \mathbb{P}^1$ and $\bar{X}$ is a natural toric compactification of $X$.
The zero section $C_0\simeq \mathbb{P}^1$ of the vector bundle $X$ intersect $D$ at one point $p$ and let us choose a planar curve $C\subset D$ containing $p$ and $\bar{C}\subset \bar{X}$ is its natural compactification.

Let $\mathcal{P}(\bar{X},\bar{C}_\lambda,r,n)$ be a moduli space of pairs of a pure one-dimensional sheaf $\mathcal{G}$ with support on $C\cup C_0$ and the map $s: \mathcal{O}_X\to \mathcal{G}$ such that 
the its cokernel has zero-dimensional support, $\ch_2(\mathcal{G})=|\lambda|[\bar{C}]+r[C_0]$ and $\chi(\mathcal{G})=\chi(C)+n$ and the kernel of $s$ is the ideal sheaf defining $\bar{C}_\lambda$. 
Respectively, the $\mathcal{P}(\bar{X},\bar{C}_\lambda,p,r,n)$ is the moduli space of pairs with the extra restriction that the cokernel is set-theoretically supported on
$C_0$.  Finally, let us denote by $\mathcal{P}(X,r,n)$ the moduli space of pairs $(\mathcal{G},s)$ of one-dimensional pure sheaves $\mathcal{G}$ on $\bar{X}$ supported on $C_0$ with  $\ch_2(\mathcal{G})=r[C_0]$ and $\chi(\mathcal{G})=\chi(C)+n$.  As explained in the introduction, $(r,n)$ can be regarded as D2-D0 charges and $\mathcal{P}(\bar{X},\bar{C}_\lambda,p,r,n)$ is physically the moduli space of D6-D2-D2'-D0 bound states. 

The topological vertex formula in the context of PT theory \cite{PT1, MOOP} implies the equivalence between the generating function for Euler characteristics of the pair moduli spaces and the large $N$ limit of Witten's quantum invariant of $S^3$:
$$Z(X,a,q):=\sum_{r,n} q^{2n}a^{2r}\chi(\mathcal{P}(X,r,n))=\prod_{i>0}\frac{1}{(1-a^2q^{2i})^i}~.$$
Using the technique of wall-crossing, the authors of \cite{DHS} prove  the equivalence between the topology of the Hilbert schemes of points on the curve and the moduli space of pairs up to the $S^3$ invariant $Z(X,a,q)$:

\begin{theorem}\label{thm:DHS}\text{\cite{DHS}}
$$ Z(X,a,q)\sum_{r,n} (-a^2)^{r} q^{2n} \chi(C_p^{[n,n+r]})=\sum_{r,n} a^{2r} q^{2n} \chi(\mathcal{P}(X,C_{\yng(1)},p,r,n)~.$$
\end{theorem}

This theorem has motivated the authors of \cite{DHS} to extend the conjecture \eqref{O-ober-eq} for colored HOMFLY invariants, which was proven in \cite{Mau}:
\begin{theorem}\text{\cite{Mau}}
\be\nonumber
Z(X,a,q) \overline P_\lambda(L_{C,(0,0)})=\sum_{r,n}a^{2r} q^{2n} \chi(\mathcal{P}(X,C_\lambda,p,r,n))~.
\ee
\end{theorem}
Note that, in the case of the unknot and the Hopf link $T_{2,2}$, the formula immediately follows from one- and two-leg PT-vertex theory respectively \cite{PT1,MOOP}.

In the following, let us state the conjecture for the homological version. Given an algebraic variety $X$, we denote by $[X]$ the corresponding element of the Grotendieck ring $K_0(Var/\mathbb{C})$ of the complex algebraic varieties. Let us assume the following conjecture:

\begin{conjecture} The classes $[\mathcal{P}(X,n,r)]$ and $[\mathcal{P}(X,C,p,n,r)]$ belong to the subring of $K_0(Var/\mathbb{C})$ generated by the class of the affine line $\mathbb{L}:=[\mathbb{A}^1]$.
\end{conjecture}

\noindent Assuming this conjecture, we state the relation between colored superpolynomial and motivic Donaldoson-Thomas invariants:

\begin{conjecture}\text{\cite{DHS}} There is a universal function $g_{\lambda}:\ZZ^2\to \ZZ$ such that 
$$\overline\scP_\lambda(L_{C,(0,0)};a,q,t_c=\bL)\sum_{r,n} a^{2r} q^{2n} [\mathcal{P}(X,r,n)]=\sum_{r,n}a^{2r} q^{2n} \mathbb{L}^{g_\lambda(r,n)} [\mathcal{P}(X,C_\lambda,p,r,n)]~.$$
\end{conjecture}

\noindent Here we assume the existence of colored HOMFLY homology and the homological grading should be consistent with $t_c$-gradings in \S\ref{sec:quadruple}.
The relation between this conjecture and Conjecture \ref{planar-homology} is explained by the motivic wall-crossing formalism. More precisely, up to a certain technical conjecture in the motivic Donaldson-Thomas theory, the paper \cite{DHS} provides a proof of the following formula
$$Z^{mot}(X,a,q,\mathbb{L})\sum_{r,n}(-a^2)^{r} q^{2n}\mathbb{L}^{r^2}~[C_p^{[n,n+r]}]=\sum_{r,n}a^{2r}  q^{2n} \mathbb{L}^{r^2}~[\mathcal{P}(X,C_{\yng(1)},p,r,n)]~.$$
where by $Z^{mot}(X,a,q,\mathbb{L}):=\sum_{r,n} a^{2r} q^{2n} [\mathcal{P}(X,r,n)]$ we denote the motivic invariant of the resolved conifold.

\section{Torus knot homology and DAHA}\label{sec:DAHA}
 In this section, we provide the definition of DAHA-superpolynomials of $\GL_N$-type, which are equivalent to refined Chern-Simons invariants \cite{AS}. For the general definition including other types, we refer to \cite{Ch,Ch2}.

Let us first define the double affine Hecke algebra $\mathcal{H\!\!\!H}_N$ of $\GL_{N}$-type \cite{CherednikNotes}. 
\begin{definition}
The algebra $\mathcal{H\!\!\!H}_N$ is defined by generators $T_i^{\pm 1}$ for  $i\in \{1,\ldots,N-1\}$ and ${X_j}^{\pm 1}, {Y_j}^{\pm 1}$ for  $j\in \{1,\ldots,N\}$, under the following relations:
\begin{multline}
\label{DAHA relations}
\hspace{2cm}(T_i+\mathfrak{t}^{-\frac12})(T_i-\mathfrak{t}^{\frac12})=0,\qquad  T_{i}T_{i+1}T_{i}=T_{i+1}T_{i}T_{i+1}, \\
T_iX_iT_i=X_{i+1},\qquad  {T_i}^{-1}Y_i{T_{i}}^{-1}=Y_{i+1}\\
[T_i,T_k]=0,\qquad [T_i,X_k]=0,\qquad [T_i,Y_k]=0,\qquad  \textrm{for}\ \ |i-k|>1\\
[X_j,X_k]=0,\qquad [Y_j,Y_k]=0, \\ Y_1X_1\ldots X_N=\mathfrak{q}^{\frac12}X_1\ldots X_NY_1,\quad
{X_1}^{-1}Y_2=Y_2{X_1}^{-1}{T_1}^{-2} \\\nonumber
\end{multline} 
\end{definition}
In fact, it contains two copies of the affine Hecke algebras generated by $(T_i,X_j)$ and $(T_i,Y_j)$. For an element $w\in W=S_N$ of the Weyl group and its representation $w=s_{i_1}\cdots s_{i_j}$ by transpositions ($s_i=(i,i+1)$), we define $T_{w} :=T_{i_1}\cdots T_{i_j}$. From the definition of $\mathcal{H\!\!\!H}_N$, this is independent of a representation $w$. In addition, for a set of non-negative integers $\lambda=(\lambda_1,\ldots,\lambda_N)$,  we define $X_\lambda:=\prod_{i=1}^{N}X_{i}^{\lambda_i}$ and $Y_\lambda:=\prod_{i=1}^{N}Y_{i}^{\lambda_i}$. Denoting a ring of Laurent polynomials of $\frakq^{\frac12},\frakt^{\frac12}$ by $\bK_0:=\bC[\frakq^{\pm\frac12},\frakt^{\pm\frac12}]$, we can state the PBW theorem for DAHA.

\begin{theorem}[PBW Theorem]\label{PBW} Any $h\in \mathcal{H\!\!\!H}_N$ can be written uniquely in the form 
$$
h =\sum_{\lambda,w,\mu}c_{\lambda,w,\mu}X_\lambda T_wY_\mu~,
$$
for $c_{\lambda,w,\mu}\in \bK_0$.
\end{theorem}

It is well-known that it admits a \emph{polynomial representation}
\bea\label{poly-rep}
p: \mathcal{H\!\!\!H}_N &\longrightarrow& \textrm{End}(\bK_0[x_1,\ldots,x_N])~.
\eea
Here $X_i$ acts as the multiplication of $x_i$ and the action of $T_i$ is given by  the Demazure-Lusztig operators
\be\nonumber
p(T_i)= \frakt^{1/2}s_{i}+(\frakt^{1/2}-\frakt^{-1/2}) \frac {s_i-1}{x_i/x_{i+1}-1}~.
\ee
In addition, the action of $Y_i$ follows from 
\bea\nonumber
Y_i= T_{i}\cdots T_{N-1}\sigma_{\pi} T_{1}^{-1}\cdots T_{i-1}^{-1}~.
\eea
where  $\sigma_{\pi}:=s_{N-1}\cdots s_{1}\partial_{1}$ with 
$$
\partial_{i}(f)=f(x_1,\ldots,x_{i-1},\frakq x_i,x_{i+1},\ldots x_N)~.
$$

Using a central idempotent in the group algebra of the Weyl group $W$
\begin{equation}\nonumber
e :=  \displaystyle\sum_{w\in W} \frac{\frakt_wT_w}{\frakt_w^2} ~.
\end{equation}
we can define the \emph{spherical DAHA} 
$\mathcal{SH\!\!\!H}_N:= e\mathcal{H\!\!\!H}_Ne\subset\mathcal{H\!\!\!H}_N$. Note that given a reduced decomposition $w = s_{i_1} \cdots s_{i_k}$ of an element $w\in W$, we define $\frakt_w := \frakt^{\frac{k}{2}}$. There is an action of $\textrm{PSL}(2,\mathbb{Z})= \langle\tau_\pm\text{ : } \tau_+\tau^{-1}_-\tau_+ = \tau^{-1}_-\tau_+\tau^{-1}_-\rangle$ generated by
\begin{equation}\nonumber
\tau_+ = \left( \begin{array}{cc}
1 & 1\\
0 & 1\end{array} \right)~, \qquad \tau_- = \left( \begin{array}{cc}
1 & 0\\
1 & 1\end{array} \right)~,
\end{equation}
on $\mathcal{H\!\!\!H}_N$, which  can be explicitly written as
\begin{equation}\nonumber
\tau_+: \begin{cases} 
      X_i\mapsto X_i Y_i  (T_{i-1}\cdots T_i)(T_{i}\cdots T_{i-1})\\
      T_i\mapsto T_i\\
      Y_i\mapsto Y_i
   \end{cases}\  \tau_-: \begin{cases} 
      X_i\mapsto X_i\\
      T_i\mapsto T_i\\
      Y_i\mapsto Y_iX_i(T_{i-1}^{-1}\cdots T_i^{-1})(T_{i}^{-1}\cdots T_{i-1}^{-1})
   \end{cases}
\end{equation}
In fact, the action preserves the restriction to the spherical DAHA $\mathcal{SH\!\!\!H}_N\subset\mathcal{H\!\!\!H}_N$. Writing an element $h\in\mathcal{H\!\!\!H}_N$ in the form of $h =\sum_{\lambda,w,\mu}c_{\lambda,w,\mu}X_\lambda T_wY_\mu$
via the PBW Theorem \ref{PBW}, we define a map $\{\cdot\}_{ev}:\mathcal{H\!\!\!H}_N\rightarrow\bK_0$ called the \emph{evaluation coinvariant} by substituting
\begin{equation}\label{evalsub}
X_i\mapsto \frakt^{-\frac{n+1-2i}{2}}\text{, }\hspace{10pt}T_i\mapsto \frakt^{\frac{1}{2}}\text{, }\hspace{10pt}Y_i\mapsto \frakt^{\frac{n+1-2i}{2}}~.
\end{equation}
Even without the use of the PBW theorem which can be rather complicated to implement, the evaluation coinvariant $\{\cdot\}_{ev}$ can be carried out by the substitution \eqref{evalsub} in the polynomial representation \eqref{poly-rep}. 

 For $f\in \bK_0[Y_1,\ldots,Y_N]^{\textrm{sym}}$, we define a Macdonald polynomial $M_\lambda(X)$ of $\GL_{N}$-type with a dominant weight $\lambda\in P_+$ by
\be\nonumber
p(f)\cdot M_\lambda(X) = f(\frakt^\rho \frakq^\lambda)M_\lambda(X)~, \quad M_\lambda(X)=m_\lambda+\sum_{\mu<\lambda} c_{\lambda,\mu} m_\mu,
\ee
where $m_\nu$ is the sum of the elements in $S_N$ orbit of $X^\nu$ and $<$ is the dominance partial order on the partitions. If we denote  a field of rational functions of $\frakq,\frakt$ by $\bK:=\bC(\frakq^{\frac12},\frakt^{\frac12})$,  and  an algebra of symmetric polynomials $\bK[x_1,\ldots,x_N]^{\textrm{sym}}$ over $\bK$ by  $V_N^{\textrm{sym}}$, then the Macdonald polynomials span an orthogonal basis of $V_N^{\textrm{sym}}$ with respect to the Macdonald pairing \eqref{mac-pair}.

Now let us define DAHA-Jones polynomials.  Indeed, $M_\lambda(X)/M_\lambda(\frakt^{-\rho})$ is an element of the spherical DAHA $\mathcal{SH\!\!\!H}_N$. Therefore, for the $(m,n)$ torus knot, 
we choose an element ${\gamma}_{m,n}\in \textrm{PSL}(2,\mathbb{Z})$
\begin{equation}\nonumber
\gamma_{m,n} = \left( \begin{array}{cc}
m & \ast\\
n & \ast\end{array} \right)~,
\end{equation}
such that \emph{reduced} DAHA-Jones polynomial is defined by
\begin{gather}\nonumber
\scP^{\textrm{DAHA}}_{\fraksl(N),\lambda}(T_{m,n}; \frakq,\frakt) := \{{\gamma}_{m,n}(M_\lambda)/M_\lambda(\frakt^{-\rho})\}_{ev}~.
\end{gather}
This is indeed a Laurent polynomial of $\frakq,\frakt$. The specialization $\frakq=\frakt=q^2$ leads to $\lambda$-colored $\fraksl(N)$ quantum invariants of the $(m,n)$ torus knot. In addition, the existence of DAHA-superpolynomials $\scP^{\textrm{DAHA}}_{\lambda}(T_{m,n};\fraka,\frakq,\frakt)$ has been proven:

\begin{theorem} [Stabilization]\text{\cite{GN}}
\label{conj1}
There exists a unique polynomial $\scP^{\textrm{DAHA}}_{\lambda}(T_{m,n};\fraka,\frakq,\frakt)$ such that:
$$
\scP^{\textrm{DAHA}}_{\fraksl(N),\lambda}(T_{m,n};\frakq,\frakt) = \scP^{\textrm{DAHA}}_{\lambda}(T_{m,n};\fraka=\frakt^N,\frakq,\frakt)~.
$$
\end{theorem}
For the DAHA-superpolynomials, the mirror/transposition symmetry (see \eqref{mirror-HOMFLY}) $$ \scP^{\textrm{DAHA}}_{\lambda^T}(T_{m,n};\fraka,\frakq,\frakt)=\scP^{\textrm{DAHA}}_{\lambda}(T_{m,n};\fraka,\frakt^{-1},\frakq^{-1}) ~,$$ and the refined exponential growth property (see  (\ref{exp-growth-HOMFLY-1},\ref{exp-growth-HOMFLY-2}))  
\bea\nonumber
 \scP^{\textrm{DAHA}}_{\sum_{i=1}^\ell \lambda_i \omega_i}(T_{m,n};\fraka,\frakq=1,\frakt)& =& \prod_{i=1}^\ell\Big[\scP^{\textrm{DAHA}}_{ \omega_i}(T_{m,n};\fraka,\frakq=1,\frakt)\Big]^{\lambda_i}~,
 \eea
  have been proven in \cite{Ch2}.

Moreover, the DAHA-Jones polynomials agree with refined Chern-Simons invariants \cite{AS}. It is well-known that there is the $\SL(2,\bZ)$ transformation on the quantum Hilbert space of Chern-Simons theory on a torus $T^2$ \cite{CSJones}, which is generated by the modular $S$ and $T$ matrices in the WZW model.
From the perspective of refined topological strings,
refined Chern-Simons theory is formulated in \cite{AS}, in which
 deformations of the modular
$S$ and $T$ matrices \cite{Kirillov} are the main constituents: 
\be\nonumber
S_{\lambda\mu}=S_{00}~ M_{\lambda}(\frakt^\rho\frakq^\mu)M_{\mu}(\frakt^\rho)~,\qquad T_{\lambda\mu}=\delta_{\lambda\mu} \frakq^{\frac12\sum_i\lambda_i(\lambda_i-1)}\frakt^{\sum_i\lambda_i(i-1)}~.
\ee
This enables the evaluation of refined torus knot
invariants by using knot operators, which are proven to be equal to the DAHA-Jones polynomials \cite{GN}. In addition, let us note that the large $N$-limits of the refined modular $S$ and $T$-matrix are expressed in terms of the Hilbert schemes of points on the affine plane $\bC^2$ \cite{Na}.

When colors are specified by rectangular Young diagrams, the DAHA-superpolynomials with the change of variables
\be\label{cov}
\fraka=-a^2 t_c ~,\quad \frakq= q^2t_c^2 ~,\quad \frakt=q^2~,
\ee
conjecturally coincide with Poincar\'e polynomial polynomials of colored HOMFLY homology of the corresponding torus knot with $t_c$-gradings. For non-rectangular Young diagrams, it is known that the DAHA-superpolynomials  include both positive and negative signs after the change of the variables \eqref{cov}.

\section{Torus knot homology, Hilbert schemes and rational  DAHA}\label{sec:rational-DAHA}

\subsection{Torus knot homology and rational DAHA }\label{subsec:torus ratDAHA}
In recent years, it has been proven that the underlying symmetry of torus knot homology is dictated by the finite dimensional representations of the rational Cherednik algebra (rational DAHA). Therefore, in this subsection, we shall explain how the finite dimensional module of the rational DAHA gives rise to torus knot invariants. In addition, we will interpret it in terms of geometric representation theory, \textit{i.e.} as an action on cohomology groups of Hilbert schemes of points on planar curves.

To begin with, let us quickly review the definition and some basic properties of the rational DAHA. The rational DAHA $\bar{\Hh}_c(S_n)$ associated to $\mathfrak{gl}(n)$ \cite{EG} is the quotient of the semi-direct product of the free algebra on 
$x_i,y_i$, $(i=1,\dots,n)$ with the symmetric group $S_n$ by the following set of relations:
\begin{gather*} 
[x_i,x_j]=[y_i,y_j]=0, \quad [x_i,y_j]=cs_{ij}\quad \textrm{for} \ \  i\ne j,\quad [x_i,y_i]=1-c\sum_{j\ne i}s_{ij},
\end{gather*}
where \(S_n\ni s_{ij}\) is the transposition that exchanges \(i\) and
\(j\).  
Indeed, the rational DAHA $\bar{\Hh}_c(S_n)$ is the universal flat deformation of the algebra $\bar{\Hh}_0(S_n)$ that is  the semi-direct product of the algebra of differential operators with 
$S_n$ \cite{EG}. Roughly speaking, the rational DAHA is a ``Lie algebra" of the DAHA
\cite{Ch} in \S \ref{sec:DAHA} where the parameter $c$ appears as $\frakt=\frakq^c$. In physics, $c$ is the parameter for the $\beta$-deformation.

The elements $x_1+\dots+x_n$ and $y_1+\dots+y_n$ generate an algebra of differential operators 
of one variable $D=\CC[x,\frac{\partial}{\partial x}]$, and we have a direct product decomposition $\bar{\Hh}_c(S_n)=\Hh_c(S_n)\otimes D$
where $\Hh_c(S_n)$ is the subalgebra generated by $S_n$ and the elements $x_i-x_j,y_i-y_j$.

The representation theory of the above Cherednik algebras is well understood \cite{BEG3}. In particular, when $c=\frac{m}{n}$ with co-prime $(m,n)$,
there exists a unique irreducible finite-dimensional representation of $\Hh_c(S_n)$. We use the notation 
$L_{\frac{m}{n}}$ and $\overline{L}_{\frac{m}{n}}$ for the irreducible finite dimensional of $\Hh_c(S_n)$ and its $\bar{\Hh}_c(S_n)$
counterpart.

In addition, as in \S \ref{sec:DAHA}, the rational DAHA also admits a \emph{polynomial representation}
\be\nonumber
p:\overline\Hh_c(S_n)\to \End(\bC[x_1,\ldots,x_n])~.
\ee
Again, $X_i$ acts as the multiplication by $x_i$ and the symmetric group \(S_n\) acts naturally as permutations of the variables \(x_i\). The action of $y_i$ is given by the \emph{Dunkl operator} $D_i$
\be\nonumber
p(y_i)=D_i:=\frac{\partial }{\partial x_i} + c \sum_{j \neq i} \frac{s_{ij} - 1}{x_i-x_j}~.
\ee
 In the setting of the highest weight categories for the rational DAHA \cite{GGOR}, the polynomial representation
is an example of the Verma type module and, in particular, it admits the unique irreducible quotient $\overline{L}_{\frac{m}{n}}$. By restriction, we can obtain a polynomial representation
 $\Hh_c(S_n)\to \End(\CC[x_1-x_2,\dots,x_{n-1}-x_n])$ and its irreducible quotient $L_{\frac{m}{n}}$.

\vspace{.3cm}
\noindent\textbf{Example:}
\label{example:n/2}
In the case of \(n=2\), we can write $u = x_1 - x_2$ so that
the unique non-trival element  $s \in S_2$ sends \(u\) to \(-u\). In addition, the Dunkl operator acts as 
$$D_1(u^m) =  m u^{m-1} +c\frac{(-u)^m - u^m}{u} = \begin{cases}
  mu^{m-1} \quad & m\  \text{even} \\  (m- 2 c) u^{m-1} \quad & m \ \text{odd} \end{cases}
$$
and \(D_2(u^m)=-D_1(u^m)\). Hence, 
we see that the Dunkl operators have a
 nontrivial kernel if and only if \(c = m/2\), where \(m\) is an odd
 integer. In this case, \(\Hh_{\frac{m}{2}}(S_2)\) has a finite dimensional
 representation of the form \(\CC[u]/(u^m)\). 
\vspace{.3cm}

Remarkably, the finite dimensional representation $\overline{L}_{\frac{m}{n}}$ of the rational DAHA turns out to be geometrically realized. Let $C_{n,m}$ be a planar curve defined by $x^m=y^n$ and let us denote by $e_i\in \CC[S_n]$ the Young projector  on irreducible $S_n$-representation
$\Lambda^i \CC^{n-1}$ \cite{Fulton-Harris}. Then, the spherical rational DAHA acts on the cohomology groups of Hilbert schemes of the curve $C_{n,m}$:
\begin{theorem}\text{\cite{OY2}}\label{action} The algebra $e_k \bar{\Hh}_{m/n}(S_n) e_k$ acts on the vector space  
 $$H^*(C^{[\bullet,\bullet+k]}_{m,n}):=\oplus_{\ell}  H^*( C_{m,n}^{[\ell,\ell+k]})$$
 by means of Hecke-Nakajima type operators, and this module is isomorphic to $e_k\overline{L}_{\frac{m}{n}}$. 
 \end{theorem}

The existence of the action of the spherical rational DAHA $e_k\bar{\Hh}_{\frac{m}{n}}(S_n)e_k$ on the cohomology $H^*(C^{[\bullet,\bullet+k]}_{m,n})$  could be extracted from the  action of the algebra $\Hh_{\frac{m}{n}}(S_n)$ on the cohomology $H^*(\mathrm{Sp}_{\frac{m}{n}})$ of 
the homogeneous affine Springer fiber studied in \cite{OY1}. Indeed, it is shown in \cite{OY1} together with the action of $\Hh_{\frac{m}{n}}(S_n)$ on $H^*(\mathrm{Sp}_{\frac{m}{n}})$ 
 that an isomorphism $H^*(\mathrm{Sp}_{\frac{m}{n}})\simeq L_{\frac{m}{n}}$ have been constructed. On the other hand, the results of  \cite{MY,MS} imply that the cohomology of the affine Grassmannian version  of the affine Springer fiber $H^*(\mathrm{Sp}_{\frac{m}{n}}^{Gr})$ are to $H^*(C^{[\bullet]})$ 
 as $eL_{\frac{m}{n}}$ to $e\overline{L}_{\frac{m}{n}}$. Hence, in \cite{OY2}, we construct the action of $e_k \overline\Hh_{\frac{m}{n}}(S_n) e_k$ on $H^*(C_{m,n}^{[\bullet,\bullet+k]})$ directly.

 The isomorphism from the theorem matches the $\bullet$-grading on $H^*(C_{m,n}^{[\bullet,\bullet+k]})$ with 
the grading on $e_k\overline L_{\frac{m}{n}}$ given by spectral decomposition   with respect to
the action 
of the element $h=\sum_i x_iy_i$  (see \cite{BEG3} for graded characters of $\overline{L}_{\frac{m}{n}}$). 
On the other hand, the homological grading on $H^*(C_{m,n}^{[\bullet,\bullet+k]})$ is related to the filtration of Jansen type on $e_k\overline{L}_{\frac{m}{n}}$, which seems to be 
new to  experts. In \cite{GORS}, an algebraic (in terms of the rational DAHA) description of the filtration  $\mathcal{F}^\bullet$ has been proposed, which coincides with cohomological filtration.  
 The main thrust of the result \cite{GORS} is that, combining with Conjecture \ref{planar-homology}, the HOMFLY homology of torus knots should be related to
 the representation theory of the rational DAHA.
\begin{conjecture}\label{conj:ratDAHAKnotHom} \text{\cite{GORS}} The finite-dimensional module $\overline{L}_{\frac{m}{n}}$ admits a filtration $\mathcal{F}^\bullet$ such that
we have an isomorphism of triply-graded spaces:
$$\overline \scH_{\yng(1)}(T_{m,n})\simeq \oplus_k \textrm{gr}_*^{\mathcal{F}}e_k \overline{L}_{\frac{m}{n}}.$$
\end{conjecture}
The conjecture suggests that the HOMFLY homology $\overline\scH_{\yng(1)}(T_{m,n})_{k,*,*}$ with $a$-degree $k$ is
a module over the spherical subalgebra $e_k\bar{\Hh}_{\frac{m}{n}}(S_n)e_k$ of the rational DAHA. Constructing such an action is probably
 a major challenge and its solution will require further development  of the theory of character sheaves. 
The conjecture was checked in all cases where the corresponding knot homology is known.

It was conjectured in \cite{GORS} that there are three equivalent descriptions of the filtration above. Here we present the most algebraic description. More precisely, we explain a construction 
of the filtration $\mathcal{F}^\bullet$ on $L_{\frac{m}{n}}$ and define $\mathcal{F}^i \overline{L}_{\frac{m}{n}}:=\sum_j (x_1+\dots+x_n)^j\mathcal{F}^{i-j} L_{\frac{m}{n}}.$
 
As we will see in the example below and in \S\ref{sec:matrix}, $L_{\frac{m}{n}}$ is a quotient of the polynomial ring $\CC[x_1-x_2,\dots,x_{n-1}-x_n]$ by an ideal. Let $\mathfrak{m}\subset L_{\frac{m}{n}}$ be the maximal ideal
$\mathfrak{m}=(x_1-x_2,\dots,x_{n-1}-x_n)$. In addition, we denote the $i$-th piece of $L_{\frac{m}{n}}$ with $h$-grading by $L_{\frac{m}{n}}(i)$. Then, we can define the following filtration
$$ \mathcal{F}_i L_{\frac{m}{n}}:=\sum_j \left(\mathfrak{m}^i \cap \oplus_{k<2j-i} L_{\frac{m}{n}}(k)\right).$$
Furthermore, we can define a pairing on $L_{\frac{m}{n}}$ by $(f,g)= \left[ f(D_i)g(x)\right](0)$, which
determines the filteration by $\mathcal{F}^i L_{\frac{m}{n}}:=(\mathcal{F}_i L_{\frac{m}{n}})^{\perp}$.

 \vspace{.3cm}
 
\noindent{\textbf{Example: $(m,2)$ torus knot}}

As we have seen in the example above, we have $L_{\frac{m}{2}}\cong \bC[u]/(u^m)$. The decomposition $L_{\frac{m}{2}}=e_0L_{\frac{m}{2}}\oplus e_1L_{\frac{m}{2}}$ with the young projectors $e_0=\frac12(1+s)$ and $e_1=\frac12(1-s)$  are spanned by even and odd powers of $u$, respectively. The $q$-grading on $L_{\frac{m}{2}}$ is given by $q(u^k) = 2k -(m-1)$. Elements of $e_0L_{\frac{m}{2}}$ have $a$-grading $m-1$ and filtration grading $(m-1)/2$, while elements of $e_1L_{\frac{m}{2}}$ have $a$-grading $m+ 1$ and filtration grading $(m -3)/2$. In this example, the $d_1$ and $d_{-1}$ differentials \eqref{dn} can be realized as the action of multiplication by $u$ and the Dunkl operator $D_1$. On the top row, they are indicated by red and blue arrows, respectively, in the figure below. (See also \cite[Figure 6.5]{DGR}.)
\begin{equation}
\begin{tikzpicture}[scale=1.3]
[place/.style={circle,draw=blue!50,fill=blue!20,thick},
transition/.style={rectangle,draw=black!50,fill=black!20,thick}]

\node(A) at (-5,0) {$1$};
\node(B) at (-3,0) {$u^2$};
\node(C) at (-1,0) {$u^4$};
\node(D) at (-4,1) {$u$};
\node(E) at (-2,1) {$u^3$};

\node at (0,0.52) {$\cdots$};

\node(F) at (1,0) {$u^{m-5}$};
\node(G) at (3,0) {$u^{m-3}$};
\node(H) at (5,0) {$u^{m-1}$};
\node(I) at (2,1) {$u^{m-4}$};
\node(J) at (4,1) {$u^{m-2}$};

\draw[blue,->] (D) -- (A);
\draw[blue,->] (E) -- (B);
\draw[blue,->] (I) -- (F);
\draw[blue,->] (J) -- (G);

\draw[red,->] (D) -- (B);
\draw[red,->] (E) -- (C);
\draw[red,->] (I) -- (G);
\draw[red,->] (J) -- (H);

\end{tikzpicture}\nonumber
\end{equation}

\noindent{\textbf{Example:}
When $m=n+1$ the filtration from above has a more elementary description in terms of the space of the double harmonics \cite{haiman}:
$$DH_n:=\CC[x_1,\dots,x_n,y_1,\dots,y_n]/I_+,$$
where $I_+$ is the ideal spanned by $S_n$-invariant polynomials of $x_i,y_i$ without a constant term. It is shown in \cite{Gor} that there is
a natural isomorphism of the $S_n$ modules  $L_{(n+1)/n}\simeq DH_n\otimes \mathrm{sign}$. It identifies the $h$-grading on $L_{(n+1)/n}$ with the grading defined by 
$\deg_h x_i=1,\deg_h y_i=-1$ and the grading associated to the filtration $\mathcal{F}$ with the grading defined by $\deg_{\mathcal{F}} x_i=2,\deg_{\mathcal{F}} y_i=0$.

\vspace{.2cm}

In fact, Conjecture \ref{conj:ratDAHAKnotHom} provides the interpretation of the $d_N$ differentials \eqref{dn} in terms of the rational DAHA. The example of the differentials in $(m,2)$ torus knots given above is merely its special case. We shall account for it in more details in \S\ref{sec:matrix}.

 At the decategorified level, the relation to the rational DAHA has been proven:

\begin{theorem}\text{\cite{GORS}}\label{quasis} For any coprime $(m,n)$, we have
 $$\sum_{i=0}^n a^{2i}\; \Tr(q^h,e_i \overline L_{\frac{m}{n}})=\bar{P}_{\yng(1)}(T_{m,n};a,q).$$
\end{theorem}
The colored extensions of this theorem are given in \cite{EGL} by constructing the corresponding modules of the rational DAHA. Happily, the construction proves an important property of \emph{unreduced} colored HOMFLY polynomials of torus knots; $\overline P_\lambda(T_{m,n};\sqrt{-1}\;a,q)$ is a polynomials  in $a$ and a power series in $q$ with non-negative  coefficients for any color $\lambda$. Moreover, it was shown in \cite{EGL} that $ (q^2;q^2)_{|\lambda|}\overline P_\lambda(T_{m,n};\sqrt{-1}\;a,q)$ is a polynomial with non-negative coefficients.

\subsection{Torus knot homology and Hilbert schemes of affine plane}\label{subsec:torus Hilb}
The connection to the rational Cherednik algebra uncovers another interpretation of torus knot homology.  This stems form  a relation between the category of $\overline \Hh_c(S_n)$-modules  with  a good filtration
and the category of coherent sheaves on the Hilbert scheme $\Hilb_n(\CC^2)$ of $n$-points on $\CC^2$ given in \cite{GS,GS2}. 
 We propose  a stronger link between these categories \cite{GORS}; 
\begin{conjecture}\label{conj;sheaf} For any coprime $(m,n)$, there exists a $\CC^*\times\CC^*$-equivariant sheaf 
$F_{m,n}$ on $(\CC^2)^{[n]}_0$ and a good (in the sense of \cite{GS2}) filtration on $ \overline{L}_{\frac{m}{n}}$ 
such that 
\begin{itemize}\setlength{\parskip}{-0.1cm}
\item $\sum_i \frakt^i \frakq^h\; \textrm{gr}_i^{\mathcal{F}} e_0 \overline{L}_{\frac{m}{n}}=\chi_{\frakq,\frakt}(H^0(F_{m,n}))$~,
\item $F_{m+n,n}=F_{m,n}\otimes \scL$,
\end{itemize}
where $\chi_{\frakq,\frakt}$ is the character of the  $\CC^*\times\CC^*$-action and
$\scL$ is the positive generator of the Picard group of the Hilbert scheme.
\end{conjecture}

Via Conjecture \ref{conj:ratDAHAKnotHom}, this indicates that the superpolynomials of torus knots can be expressed by the $\bC^*\times \bC^*$-equivariant character of a sheaf $F_{m,n}$. Since the $\bC^*\times \bC^*$-equivariant character of the sheaf $F_{m,n}$ is supposed to be a $\U(1)$ instanton partition function with a codimension two defect associated to the torus knot $T_{m,n}$, as we illustrate in the introduction, the conjecture above sets up the formulation of the geometric engineering \cite{GeomEng}. When $m=1+kn$ the conjecture
holds \cite{GS2} with $F_{m,n}=\mathcal{O}_{Z}\otimes \scL^k$ where $Z$ is the punctual Hilbert scheme of points
consisting of ideals $I$ such that the support of $\mathcal{O}/I$ is $(0,0)$. For general $(m,n)$, a conjectural construction of the sheaf $F_{m,n}$
(more precisely of the complex of sheaves) is presented in \cite{GN}. Below we explain the geometric meaning of the main formula (\ref{eqn:for}) in \cite{GN}.

\subsection*{DAHA-superpolynomial and Hilbert scheme of points $\Hilb_n(\CC^2)$}

It was known for a long time that theory of Macdonald polynomials and the geometry of the Hilbert schemes of points on the affine plane are intimately related \cite{haimpolygraph,haiman2,haimlectures}.  Since DAHA provides an algebraic 
setting for the theory of Macdonald polynomials, it is natural to expect some relations between the DAHA and the Hilbert schemes of points on the plane. The most complete picture for the relation is revealed in
\cite{SV1,SV2} (see also paper \cite{FT} for some results in the same direction). Using the results in \cite{SV1,SV2}, the authors of \cite{GN} have constructed the sheaf that is conjecturally equivalent to $F_{m,n}$ in Conjecture \ref{conj;sheaf} and have found the relation to the DAHA-superpolynomials.

As we see in \S \ref{sec:DAHA},  the DAHA-Jones polynomials have been constructed by using the $\textrm{PSL}(2,\bZ)$ action on a spherical subalgebra $\mathcal{SH\!\!\!H}_N$ of DAHA $\mathcal{H\!\!\!H}_N$. It is shown in \cite{SV1} that there exists an algebra $\mathcal{SH\!\!\!H}_\infty$
that admits surjective homomorphisms $\Psi_N: \mathcal{SH\!\!\!H}_{\infty}\to \mathcal{SH\!\!\!H}_N$. The algebra $\mathcal{SH\!\!\!H}_\infty$ is indeed an inductive limit of $\mathcal{SH\!\!\!H}_N$, and the restriction of
the map on the commutative subalgebra of symmetric polynomials of $X_i$ is the usual restriction map $V_\infty^{\textrm{sym}}\to V_N^{\textrm{sym}}$. It is also shown in \cite{BS} that the algebra
$\mathcal{SH\!\!\!H}_\infty$ has a natural $\textrm{PSL}(2,\ZZ)$-action and $\Psi_N$ intertwines the $\textrm{PSL}(2,\ZZ)$ actions on $\mathcal{SH\!\!\!H}_\infty$ and that on $\mathcal{SH\!\!\!H}_N$.
Thus, there is a well defined element of $\gamma_{m,n}(M_\lambda)\in S\mathcal{H\!\! \!H}_\infty$ where $M_\lambda$ represents a Macdonald polynomial with infinitely many variables. In addition, it is not hard to see that there is an evaluation map 
$\{\cdot\}_{ev,\infty}: S\mathcal{H\!\!\! H_\infty}\to\mathbb{K}_0(\fraka)$ such that  $\Psi_N$ intertwines $\{\cdot\}_{ev,\infty}|_{\fraka=\mathfrak{t}^N}$  and 
the evaluation map on $\mathcal{H\!\!\! H}_N$.
 This observation immediately implies the stabilization statement (see Theorem~\ref{conj1} and \cite{GN} for more details).

 Using known properties of the Macdonald polynomials,
  the unreduced DAHA-Jones polynomial has been constructed in terms of Macdonald pairing \cite{GN}:
 \begin{equation}\label{eqn:DAHAformula}
\overline \scP^{\textrm{DAHA}}_{\mathfrak{sl}(N),\lambda}(T_{m,n},\mathfrak{q},\mathfrak{t})=\langle v(\mathfrak{a})|\gamma_{m,n}(M_\lambda)|1\rangle_{\frakq,\frakt}\Big|_{\fraka=\mathfrak{t}^N},
 \end{equation}
 where $\langle\cdot|\cdot\rangle_{\frakq,\frakt}$ is the Macdonald pairing on  $V_N^{\textrm{sym}}$ and 
\bea\nonumber
v(\mathfrak{a})=\sum_{\lambda} \frac{1}{h_\lambda h'_\lambda}\frac{M_\lambda(X)}{M_\lambda(\mathfrak{t}^\rho)}\prod (\mathfrak{t}^{l'(\yng(1))}-\mathfrak{a}\frakq^{a'(\yng(1))})^2\in V_N^{\textrm{sym}}\otimes \mathbb{C}[\fraka]~,\cr
 h_\lambda=\prod_{\yng(1)\in \lambda} (1-\frakq^{a(\yng(1))} \frakt^{\ell(\yng(1))+1})~,  \quad h'_\lambda=\prod_{\yng(1)\in\lambda}(1-\frakq^{a(\yng(1))+1}\frakt^{\ell(\yng(1))})~.
 \eea
By taking the stable limit as $N$ goes to infinity in the  formula (\ref{eqn:DAHAformula}), we obtain a formula for the DAHA-superpolynomial in terms of the natural action of $\mathcal{SH\!\!\!H}_\infty$ on $V_\infty^{\textrm{sym}}$ and
the Macdonald pairing on $V_\infty^{\textrm{sym}}$ which is given by
\bea\label{mac-pair}
\langle { M_\lambda },{ M_\mu }\rangle_{\frakq,\frakt}=\delta_{\lambda,\mu}\prod_{(i,j)\in\lambda}
\frac{1-\frakq^{\lambda_i-j+1} \frakt^{\lambda^T_j-i}}{ 1-\frakq^{\lambda_i-j  } \frakt^{\lambda^T_j-i+1}}~.
\eea
The projection $V^{\textrm{sym}}_\infty\to V^{\textrm{sym}}_N$ is an orthogonal projection that annihilates the Macdonald polynomials $M_\lambda$ with 
$\lambda=\lambda_1\ge \dots \ge\lambda_M>0$ with $M>N$ and preserves the norm of the rest of the Macdonald polynomials.
 
 Interestingly, there is a natural identification \cite{haimlectures} 
 $$V_\infty^{\textrm{sym}} \cong \oplus_d K_{\CC^*\times\CC^*}(\Hilb_d(\CC^2))~,$$ where $K_{\CC^*\times\CC^*} (\Hilb_d(\CC^2))$ stands for 
the equivariant $K$-theory of the coherent sheaves on $\Hilb_d(\CC^2)$. This identification transfers the Macdonald pairing to the natural pairing between the equivariant coherent sheaves:
$\langle\mathcal{F},\mathcal{G}\rangle=\chi_{\mathfrak{q},\mathfrak{t}}(\mathcal{F}\otimes\mathcal{G})$ and the Macdonald polynomial $M_\lambda$ gets identified with a sky-scraper $I_\lambda$ sheaf supported on the monomial ideal 
$\mathcal{I}_\lambda=(x^{\lambda_1}y^0,x^{\lambda_2}y^1,\dots)$ with 
the profile $\lambda$.

Based on this identification, a geometric construction for the $\textrm{PSL}(2,\bZ)$ action of $\mathcal{SH\!\!\!H}_\infty$ on $\oplus_d K_{\CC^*\times\CC^*}(\Hilb_d(\CC^2))$ is presented in \cite{SV2} (see also \cite{FT} for the shuffle algebra version
of the action and \cite{Neg2} for an explanation why these actions coincide). This action was further explored in \cite{Neg}, in particular,  provides and explicit formula for the matrix coefficients  of $\oplus_d K_{\CC^*\times\CC^*}(\Hilb_d(\CC^2))$ in the basis $\{  I_\lambda\}$.
Theorem \ref{thm:main} in the introduction is obtained by combining the matrix coefficients  of $\oplus_d K_{\CC^*\times\CC^*}(\Hilb_d(\CC^2))$, the formula (\ref{eqn:DAHAformula}) and the localization of equivariant Grothendieck-Riemann-Roch formula
(see details in \cite{GN}).

The constructions of \cite{Neg} stem from the study of the nested Hilbert scheme $$\Hilb_{d,d+k}(\CC^2)\subset \Hilb_{d}(\CC^2)\times\dots\times \Hilb_{d+k}(\CC^2)$$ consisting of chains of ideals $\mathcal{I}_d\supset
\dots\supset\mathcal{I}_{d+k}$. This variety is singular but it is expected to be a local complete intersection. To avoid technical problems, the paper \cite{Neg} treats this scheme as DG scheme.
The space $\Hilb_{d,d+k}(\CC^2)$ carries natural line bundles $\mathcal{L}_i$ where the fiber of $\mathcal{L}_i$ over the chain  $\mathcal{I}_d\supset
\dots\supset\mathcal{I}_{d+k}$ is the quotient $\mathcal{I}_{i}/\mathcal{I}_{i+1}$. The geometric description of the operator $\gamma_{m,n}(M_\lambda)$  from \cite{Neg} involves push-forwards and pull-backs along the natural 
maps $$p_+: \Hilb_{d,d+k}\to \Hilb_{d,d+k-1}~,\quad p_-: \Hilb_{d,d+k}\to \Hilb_{d+1,d+k}$$ combined with twisting by the line bundles $\mathcal{L}_i$. The details of the construction are beautifully explained in the original paper, and in these notes we just state the simplest corollary of these geometric constructions.

\begin{theorem}\text{\cite{GN}}\label{fmn-GN} Let $S_{\frac{m}{n}}(i)$ be the one defined in \eqref{def:smn}. Then we have
$$\overline\scP_{\yng(1)}^{\textrm{DAHA}}(T_{m,n};\mathfrak{a}=0,\mathfrak{q},\mathfrak{t})=\chi_{\mathfrak{q},\mathfrak{t}}(F_{m,n}), \qquad F_{m,n}:= p_*(\mathcal{L}^{S_{\frac{m}{n}}(1)}_1\otimes\dots\otimes \mathcal{L}_n^{S_{\frac{m}{n}}(n)})$$
where $p$ is the natural projection $\Hilb_{0,n}(\CC^2)\to\Hilb_{n}(\CC^2).$
\end{theorem}

The last theorem combined with Conjecture~\ref{conj;sheaf} results in the formula for the character of 
$e\overline L_{\frac{m}{n}}$ in terms of some particular sheaf on the $\Hilb_n(\CC^2)$. The relation between the sheaves on $\Hilb_n(\CC^2)$ and the characters of the modules over the rational DAHA was previously 
observed in \cite{GS2}. One of the result of \cite{GS2} is the functor $\Phi_c$ from the category $\mathcal{O}$ of modules of the rational DAHA of type $A_{n-1}$ to the category of coherent sheaves on $\Hilb_n(\CC^2)$.
In particular, the paper \cite{GS2} shows that $\Phi_{(nk+1)/n}(L_{(nk+1)/n})=\mathcal{O}_Z\otimes \scL^k$. However for the other values of $m,n$ the techniques of \cite{GS2} fail to produce a description of the sheaf $\Phi_{\frac{m}{n}}(L_{\frac{m}{n}})$. Nevertheless, the sheaves $F_{m,n}$ in Theorem \ref{fmn-GN} behave as predicted by Conjecture~\ref{conj;sheaf}: $F_{1+kn,n}=\mathcal{O}_Z\otimes \scL^k$;
$F_{m+n,n}=F_{m,n}\otimes\scL$ because of $\mathcal{L}_1\otimes\dots\otimes \mathcal{L}_n=p^*(\scL)$. Thus we expect that these sheaves are the ones that are 
given by the functor from \cite{GS2}.

Let us conclude this section by mentioning the implication to physics. The character $\chi_{\frakq,\frakt}(F_{m,n})$ is supposed to be the $\U(1)$ instanton partition function with codimension two defect associated to a torus knot $T_{m,n}$ in the $\Omega$-background. The codimension two defect should be realized either as imposing the singular behavior to elementary fields on the support of the defect  or as coupling 5d $\cN=1$ $\U(1)$ gauge theory to 3d $\cN=2$ gauge theory like a surface operator of Gukov-Witten type \cite{surfaceop}. In addition, the DAHA is expected to arise as an algebra of line operators in the presence of the defect in the 5d theory. We wish that these physical properties of the defect associated to a torus knot will be understood more precisely in near future.

\section{Matrix factorizations and Koszul models}\label{sec:matrix}

Originally, knot homology has been defined by using a category of matrix factorizations \cite{KR1,KR2,Wu,Yo}. In this formulation, one ``resolves'' crossings by replacing $\raisebox{-.1cm}{\includegraphics[width=.5cm]{overcrossing}}$ and $\raisebox{-.1cm}{\includegraphics[width=.5cm]{undercrossing}}$ by two local diagrams $\raisebox{-.1cm}{\includegraphics[width=.5cm]{smoothing}}$ and $\raisebox{-.1cm}{\includegraphics[width=.4cm]{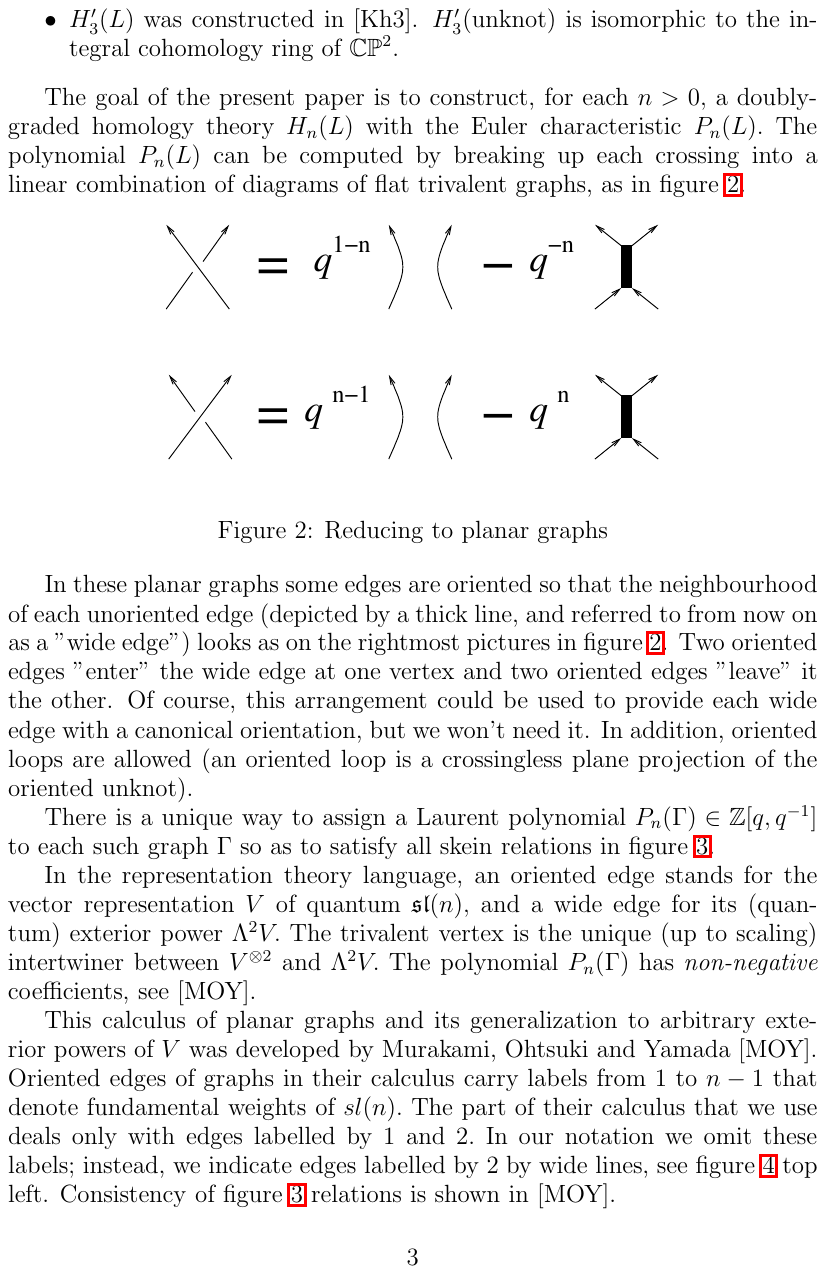}}$, which gives a cube of Murakami-Ohtsuki-Yamada graphs \cite{MOY}. Then, one assigns a matrix factorization specified by a potential $W_{\frakg,\lambda}$ for each resolution and one subsequently obtains a mapping cone of two matrix factorizations at each crossing of a knot. Then, taking a tensor product of matrix factorizations for a MOY graph, one obtains double complex whose homology is $(\frakg,\lambda)$ homology of a knot. The computations in the uncolored case have been carried out for knots with a few crossing in \cite{CM}.

Physically, this Khovanov-Rozansky setup is described by 2d Landau-Ginzburg model with topological defects on the $\bR_t\times K$ in the M-theory set-up \eqref{deformed}  \cite{GNSS}. In particular, in the case of the unknot, there is no topological defect so that the $(\frakg,\lambda)$ homology of the unknot is isomorphic to the Jacobi ring of the potential in the LG model
\be\nonumber
\scH_{\frak{g}, \lambda} (\unknot) \; = \; \textrm{Jac} (W_{\frak{g}, \lambda}) \,.
\ee
Interestingly, the brane configuration in \eqref{deformed} contains important information about the potential $W_{\frakg,\lambda}$ of the LG model. As in \cite[Figure 2]{DSV}, as the spectator M5'-brane is pushed to the fiber direction of $T^*S^3$, an M2'-brane is dynamically generated. Therefore, the left and right configurations in \eqref{deformed} are indeed related via the Hanany-Witten effect. The low-energy effective theory of this brane configuration is described by 2d $\cN=(2,2)$ theory with $\U(1)$ gauge group and $N$ fundamental chirals. The Higgs branch of the theory is $\bP^{N-1}$ and its cohomology is isomorphic to the Jacobi ring of the potential $W_{\fraksl(N),\yng(1)}=x^{N+1}$.

This can be easily generalized to an $r$-rank anti-symmetric representation $(1^r)$ of $\fraksl(N)$ homology. To realize $r$-rank anti-symmetric representation $(1^r)$, we increase the number of the spectator M5'-branes (M2'-branes) to $r$. As a result, the same procedure leads to 2d $\U(r)$ gauge theory with $\U(N)$ flavor symmetry whose Higgs branch is a Grassmannian $\textrm{Gr}(r,N)$. The potential of the corresponding Landau-Ginzburg model is a homogeneous polynomial of degree $N+1$,
\be\nonumber
W_{\fraksl(N),(1^r)} \; = \; x^{N+1}_1 + \ldots + x_r^{N+1} \,.
\ee
Using the variables $u_i$ of degree $\deg (u_i) = i$, $i=1, \ldots, r$,
which are the elementary symmetric polynomials in the $x_j$,
$$
u_i \; = \; \sum_{j_1<j_2<\cdots<j_i} x_{j_1}x_{j_2} \cdots x_{j_i} \,.
$$
the potential can be organized into a generating function \cite{GW}
\be\nonumber
\sum_N (-1)^{N} t^{N+1} W_{\fraksl(N),(1^r)} (u_1, \ldots, u_r) \; = \; \log ( 1 + \sum_{i=1}^r t^i u_i) \,.
\ee
In a similar manner, the explicit form of the potentials for symmetric representations can be also expressed through a generating function \cite{GW}
\be\label{sym-potential}
\sum_{N=0}^{\infty}t^{N+r}(-1)^{N}W_{\fraksl(N),(r)}(u_1,\ldots,u_r)=\left(1+\sum_{i=1}^{r}t^{i}u_{i}\right)\ln\left(1+\sum_{i=1}^{r}t^{i}u_{i}\right)~.
\ee
For more general representations, the corresponding Higgs branch is expected to be a partial flag variety and to find the LG potentials is still an open problem.

\vspace{.5cm}

In recent years, motivated by the relation between rational Cherednik algebra and torus knot homology, a scheme is constructed in such a way that the space of its differential forms is conjectured to be isomorphic to torus knot homology \cite{GORS,GGS}. Equivalently, the torus knot homology can be written as Jacobi ring of a certain super-Landau-Ginzburg potential $W_{\textrm{super}} (K; r)$
\be\nonumber
\overline \scH_{(r)} (K) \; \cong \; \textrm{Jac} (W_{\textrm{super}} (K;r))~.
\ee
Therefore, in what follows, we give explicit constructions of Koszul complex as well as its reformulation in terms of super-Landau-Ginzburg model.

\begin{definition}
The {\em unreduced} moduli space $\overline{\cM}_{m,n}(r)$ is defined in the affine space
with the coordinates $u_1, \ldots, u_{mr}; v_1,\ldots,v_{nr}$ by the equation
\begin{equation}
\label{equv}
(1+u_1z+u_2z^2+\ldots+u_{mr}z^{mr})^{n}=(1+v_1z+v_2z^2+\ldots+v_{nr}z^{nr})^{m}~,
\end{equation}
which should hold at every coefficient of its expansion in powers of  $z$.
\end{definition}

\begin{conjecture}\text{\cite{GGS}}
\label{symmetric torus unreduced}
The {\em unreduced} $(r)$-colored homology of the $(m,n)$ torus knot is isomorphic to the space of differential forms on $\overline{\cM}_{m,n}(r)$:
\begin{equation}
\overline{\scH}_{(r)} (T_{m,n}) \; \cong \; \Omega^{\bullet}(\overline{\cM}_{m,n}(r)) \,.
\nonumber
\end{equation}
\end{conjecture}

Let us describe the gradings on the space $\Omega^{*}(\cM_{m,n}(r))$.
The $a$-grading is defined by the degree of a differential form, so that $a(u_i)=0,\ a(du_i)=2.$ Furthermore, we define the $(q,t_r,t_c)$ gradings by the formula
\bea
\label{deg ui for symmetric}
(q,t_c,t_r)[u_i]&=&\left(2i,2i-2,2\left\lfloor\frac{i-1}{r}\right\rfloor\right),\cr
(q,t_c,t_r)[du_i]&=&\left(2i-2,2i-1,2\left\lfloor\frac{i-1}{r}\right\rfloor+1\right).
\eea
It is easy to check that the defining equations of $\cM_{m,n}(r)$ are homogeneous in $q$-grading and not homogeneous in $t$-gradings. Therefore, strictly speaking, on $\Omega^{*}(\cM_{m,n}(r))$ we get  $(a,q)$ bi-grading and a pair of filtrations $(t_r,t_c)$.
\vspace{.5cm}

\noindent\textbf{Example: Unknot} We can choose $m=n=1$, so (\ref{equv}) takes the form
$$1+u_1z+u_2z^2+\ldots+u_{r}z^r=1+v_1z+v_2z^2+\ldots+v_{r}z^r~.$$
Therefore, we have $\overline{\cM}_{1,1}(r)=\Spec\bC[u_1,\ldots,u_r],$
and 
\be\label{unknot-homology}
\Omega^{\bullet}(\overline{\cM}_{1,1}(r))=\Lambda^*(\xi_1,\ldots,\xi_{r})\otimes \bC[u_1,\ldots,u_{r}]~,
\ee
where we rename $du_i$ by $\xi_i$. This agrees with the description of the unreduced $(r)$-colored triply-graded homology of the unknot with $t_c$-gradings. In the case of the fundamental representation, the diagram is drawn in Figure \ref{fig:unknot}.

\begin{figure}[H]
\centering
\includegraphics[width=10cm]{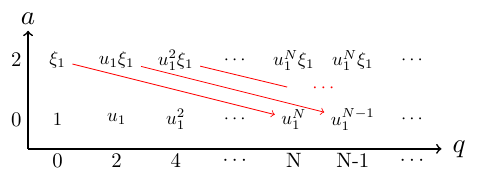}
\caption{Koszul model for unreduced uncolored HOMFLY homology $\overline \scH_{(1)}(\bigcirc)$ of the unknot. The red arrows represent the $d_N$ differential where $d(\xi_1)=\partial W_{\fraksl(N),(1)}/\partial u_1=u_1^N$ with $W_{\fraksl(N),(1)}=u_1^{N+1}$.}
\label{fig:unknot}
\end{figure}

Now let us explain how one can construct the $d_N$ differentials in this algebraic model \cite{G,G1,GOR,GORS,GGS}. First of all, it has been proven in  \cite{Stosic} that there exists the stable limit of unreduced uncolored HOMFLY homology of an $m$-strand torus knot 
$$\overline\scH_{\yng(1)}(T_{m,\infty})=\lim_{n\to \infty}(a^{-1}q))^{(m-1)(n-1)}\overline\scH_{\yng(1)}(T_{m,n})~,$$  and it is conjectured \cite[\S6]{DGR} to be isomorphic to the $(m)$-colored HOMFLY homology of the unknot
 $$\overline\scH_{\yng(1)}(T_{m,\infty}) \cong \Lambda^*(\xi_1,\ldots,\xi_{m})\otimes \bC[u_1,\ldots,u_{m}]~.$$
In addition, we use the following important theorems:
\begin{theorem}
\label{factsmf}
(a) \text{\cite{hkr}} The Hochschild cohomology of the algebra of functions on $\mathbb{C}^{n}$  is equal to the 
algebra of polyvector fields on $\mathbb{C}^{n}$.

(b) \text{\cite{dyckerhoff}} Consider a function $W:\mathbb{C}^{n}\rightarrow \mathbb{C}.$ Then Hochschild cohomology $\HHH(\MF(W))$ of the category of
matrix factorizations of $W$ is  equal to  the Koszul homology of the complex obtained from polyvector fields by the contraction with $dW$: 
\be\nonumber
\HHH(\MF(\bC[u_i],W))\cong H_*(\Lambda^*(du_i)\otimes \bC[u_i];D_W)~.
\ee
where the differential is defined by $D_{W}(du_{i})=\frac{\partial W}{ \partial u_i}$ and $D_{W}(du_{i})=0$. 

(c) \text{\cite{dyckerhoff}} If $W$ has an isolated singularity, then $\HHH(\MF(W))$  is isomorphic to the Milnor algebra of $W$ at this singularity:
$$\HHH(\MF(\bC[u_i],W))=\mathbb{C}[u_1,\ldots,u_m]/\left(\frac{\partial W}{\partial u_1},\ldots,\frac{\partial W}{ \partial u_m}\right).$$
\end{theorem}
Because the unknot homology is the Jacobi ring of the Landau-Ginzburg potential, this theorem allows us to define the $d_N$ differentials on \eqref{unknot-homology} by using the potential defined in \eqref{sym-potential}. The $d_N$ differentials in the stable homology $\scH_{\yng(1)}(T_{m,\infty})$ stay as they are in the torus knot homology $\scH_{\yng(1)}(T_{m,n})$. Hence, one can write the explicit actions of the $d_N$ differentials for $N>0$ as
\be\label{dN-koszul}
d_{N}(\xi_i) = \sum_{ j_1+\cdots+j_N=N+i-1 } u_{j_1}\ldots u_{j_N}~, \qquad d_N(u_i)=0~.
\ee
For $N\le0$, we have $d_N(u_i)=0$ as well as 
\bea\nonumber
d_{N}(\xi_i) =\delta_{N+i-1,0} \quad \textrm{for} \ N<0~,\qquad d_{N}(\xi_i) =u_{N+i-1,0} \quad \textrm{for} \ N=0~.
\eea
In a similar way, one can construct the colored differentials \cite{GGS}
\bea\label{dcolor-koszul}
d^+_{(r)\to (k)}(\xi_i)=u_{i-k}~, \qquad d^-_{(r)\to (k)}(\xi_i)=\delta_{i,r+k+1}~.
\eea

The algebraic model for reduced $(r)$-colored HOMFLY homology easily follows by setting $u_1=\cdots=u_{r}$ in the above set-up.
\begin{definition}
The {\em reduced} moduli space $\cM_{m,n}(r)$ is defined in the affine space
with the coordinates $u_{r+1},\ldots,u_{mr}; v_{r+1},\ldots,v_{nr}$ by all coefficients in the $z$-expansion of the equation
\begin{equation}
\label{equvred}
(1+u_{r+1}z^{r+1}+u_{r+2}z^{r+2}+\ldots+u_{mr}z^{mr})^{n}=(1+v_{r+1}z^{r+1}+v_{r+2}z^{r+2}+\ldots+v_{nr}z^{nr})^{m}.
\end{equation}
\end{definition}

\begin{conjecture}\text{\cite{GGS}}
\label{symmetric torus reduced}
The {\em reduced} $(r)$-colored HOMFLY homology of the $(m,n)$ torus knot is isomorphic to the space of differential forms on $\cM_{m,n}(r)$:
\begin{equation}\nonumber
\scH_{(r)} (T_{m,n}) \; \cong \; \Omega^{\bullet}(\cM_{m,n}(r)) \,.
\end{equation}
\end{conjecture}

In fact, the main result of \cite{G1}  is an explicit isomorphism of graded vector spaces
\be\nonumber
 \textrm{gr}_*^{\mathcal{F}}e_k \overline{L}_{\frac{m}{n}}\cong \Omega^{k}(\cM_{m,n}(r=1)) ~.
\ee
Thus, at $r=1$, Conjecture \ref{symmetric torus reduced} is equivalent to Conjecture \ref{conj:ratDAHAKnotHom} (up to the filtration).
Moreover, it follows from \cite[Theorem 1.4]{EGL} (colored extensions of \eqref{quasis}) that the $(a,q)$-character
of $\Omega^{*}(\overline{\cM}_{m,n}(r))$ is equal to the unreduced $(r)$-colored HOMFLY polynomial of the $(m,n)$ torus knot.

\vspace{.5cm}

\noindent\textbf{Example: Trefoil}

For example, in the uncolored case, the defining equation is 
\bea\nonumber
(1+u_2z^2)^3=(1+v_2z^2+v_3z^3)^2 \quad \longrightarrow \quad 3u_2=2v_2, \ v_3=0, \ 3u_2^2=2v_2^2~.
\eea
Thus, the space of differential form is generated by $u_2$ and $du_2$ with constraints $u_2^2=0$. Reading the degrees from \eqref{deg ui for symmetric}, one can place the generators as in Figure \ref{fig:trefoil} and its Poincar\'e polynomial is equal to \eqref{31-1}.

In a similar fashion, the reduced $(2)$-colored homology of the trefoil knot
has two even generators $u_3$ and $u_4$ and the defining equations can be obtained from (\ref{equvred}):
\be\nonumber
u_3u_4=u_4^2=u_3^3=0.
\ee
Their differentials have the form
$$u_3du_4+u_4du_3=2u_4du_4=3u_3^2du_3=0.$$
Therefore, one can check that the monomial basis in $\Omega^{*}(\cM_{2,3} (2))$ is given in Figure  \ref{fig:trefoil} and the Poincar\'e polynomial from \eqref{deg ui for symmetric} agrees with \eqref{quad31} up to the factor $a^{4}q^{-4}$.
The actions of the colored differentials $d^{\pm}_{(2)\to(1)}$ can be easily read from \eqref{dcolor-koszul} so that $d^{+}_{(2)\to(1)}(du_3)=u_3$ (red arrow) and $d^-_{(2)\to(1)}(du_i)=\delta_{i,4}$ (blue arrow). We leave it as an exercise to see the canceling differentials $d^\pm_{(2)\to(0)}$.

\begin{figure}[H]
\centering
\includegraphics[width=\textwidth]{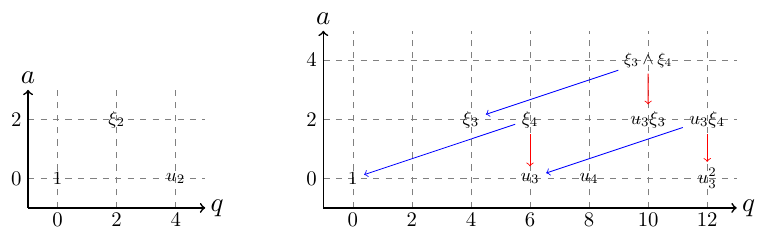}
\caption{Koszul model for reduced uncolored (left) and $(2)$-colored (right) HOMFLY homology of the trefoil. The red and blue arrows represent the $d^+_{(2)\to(1)}$ and $d^-_{(2)\to(1)}$ differentials, respectively where $d^+_{(2)\to(1)}(\xi_3)=u_3$ and  $d^-_{(2)\to(1)}(\xi_4)=1$.}
\label{fig:trefoil}
\end{figure}

\noindent\textbf{Example: $(3,4)$ torus knot }

Let us look at the case of  $(3,4)$ torus knot. Solving \eqref{equvred} with $(m,n)=(3,4)$,  the reduced uncolored homology has two bosonic generators $u_2$ and $u_3$,
with the following defining equations for $\cM_{3,4} (1)$:
\be\nonumber
u_2u_3=0~,\qquad \frac{2 u_2^3}{9} = u_3^2~.
\ee
Because the HOMFLY homology of the $(3,4)$ torus knot is thick, \textit{i.e.} the $\delta$-gradings of all the generators are not the same, there is a non-trivial action of the $d_2$ differential $d_2(du_3)=u_2^2$ (Figure \ref{fig:T34}), which can be read off from \eqref{dN-koszul}.

\begin{figure}[h]
\centering
\includegraphics[width=10cm]{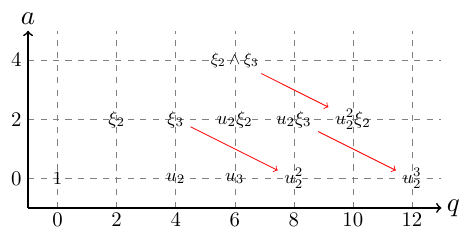}
\caption{Koszul model for reduced uncolored HOMFLY homology $\scH(T_{3,4})$ of the (3,4) torus knot. The red arrows represent the $d_2$ differential where $d_2(\xi_3)=u_2^2$.}
\label{fig:T34}
\end{figure}

\vspace{.5cm}

Now we can reformulate Conjecture \ref{symmetric torus unreduced} 
in terms of the super-Landau-Ginzburg model with a potential $W_{\textrm{super}} (K;r)$.
Let us define the potential $W (T_{m,n}; r)$ on the $u$-space given by the formula
\be
W (T_{m,n}; r) =\textrm{Coef}_{(m+n)r+1}(1+u_1z+\ldots+u_{mr}z^{mr})^{\frac{m+n}{n}}~.
\label{WTmnr}
\ee
Then, the super-Landau-Ginzburg potential can be constructed as
\be\nonumber
W_{\textrm{super}} (T_{m,n}; r) :=\sum_{i}\frac{\partial W(T_{m,n}; r) }{\partial u_{i}}\xi_{i} \,,
\ee
so that the space of differential forms $\Omega^{\bullet}\left(\overline{\cM}_{m,n}(r)\right)$ is isomorphic to the Jacobi ring of $W_{\textrm{super}} (T_{m,n}; r)$:
\begin{equation}
\label{W super}
\Omega\left[\overline{\cM}_{m,n}(r)\right]=\bC[u_1,\ldots,u_{nr},\xi_1,\ldots,\xi_{nr}]/\left(\frac{\partial W_{\textrm{super}}}{\partial u_{i}},\frac{\partial W_{\textrm{super}}}{\partial \xi_{i}}\right).
\end{equation}

To obtain the model for reduced $(r)$-colored HOMFLY homology, we set $u_1=\cdots=u_{r}$ in \eqref{W super}.
For instance, we find the following potentials from \eqref{WTmnr} with $u_1=0$ in the simple examples considered earlier:
\bea\nonumber
W_{\textrm{super}} (T_{2,3}; \yng(1))&=&3u_2^2\xi_2~,\cr
W_{\textrm{super}} (T_{3,4}; \yng(1)) &=& \left(-\frac{28 u_2^3}{243} + \frac{14 u_3^2}{27}\right) \xi_2+\frac{28 u_2 u_3}{27} \xi_3~.
\eea
It is easy to check that their Jacobi rings are isomorphic to the spaces of the differential forms in the Koszul models above.

\section{Other directions}\label{sec:other}

Although we have just reviewed some part of the recent developments on knot homology, there are many approaches which are not covered in this note. Let us list some of them.

\subsection*{Kapustin-Witten equations} 

In \cite{fiveknots}, a candidate for $\fraksl(N)$ homology of  knots is constructed based on counting the solutions of certain elliptic partial differential equations in four and five dimensions. If one takes the reduction of the circle of the cigar in the right setup of \eqref{deformed}, then we have the following brane configuration:
\be\nonumber
\begin{matrix}
{\mbox{\rm space-time:}} & \qquad & \bR_t & \times & \bR^2 & \times &\bR_+& \times  &T^* S^3 \\
{\mbox{\rm D6-brane:}} & \qquad & \bR_t &\times& &\times & & \times &T^* S^3  \\
{\mbox{\rm $N$ D4-branes:}} & \qquad & \bR_t &\times& &\times &\bR_+ & \times &  S^3  \\
{\mbox{\rm D2'-brane:}} & \qquad&  \bR_t &\times& &\times & &\times & C_K  
\end{matrix}
\ee
where we use $y\ge0$ for the coordinate of $\bR_+$. Then, Witten has considered the BPS equations of 5d $\SU(N)$ twisted super-Yang-Mills theory on $N$ D4-branes $M_5=\bR_t  \times \bR_+ \times S^3$ in the presence of the surface operator supported on $\bR_t\times K$ due to the D2'-brane:
\bea\label{KW}
F^+ - \frac{1 }{ 4} B \times B - \frac{1 }{ 2} D_y B & =& 0  ~, \cr
F_{y i} + D^{j} B_{j i} & =& 0 ~,
\eea
with the  elliptic boundary condition at $y=0$
$$
B \simeq \frac{B_0}{ y} \quad \text{as} \quad y \to 0 \,.
$$
Here, $B$ is a section of $\Omega^{2,+} (\bR_t \times S^3) \otimes {\rm ad} (E)$ with a $\SU(N)$-bundle $E\to M_5$, so that
$$
(B \times B)_{ij} = \sum_{k} [B_{ik} , B_{jk}]~.
$$
The set $\mathcal{V}$ of the solutions to the equation \eqref{KW} is bi-graded, where the homological grading is the fermion number, and  the $q$-grading is the instanton number obtained by integrating over $S^3\times \bR_+$. In addition, a supercharge $Q$ of the topological theory on $M_5$ acts the space $\mathcal{V}$ so that one can take the cohomology $\scH(K)$ of $Q$. Indeed, the vector space $\scH(K)$ is a topological invariant and the candidate for the $\fraksl(N)$ homology of a knot.

\subsection*{3d/3d correspondence} 

The effective 3d $\cN=2$ theory on $\bR_t  \times  D$ in \eqref{deformed} is what is denoted by  $T_{\fraksl(N)}[S^3\backslash K]$, which appears in the 3d/3d correspondence \cite{Dimofte:2010tz,Terashima:2011qi,DGG13,DGG14,Dimofte:2014zga}\footnote{We also refer the reader to the review \cite{Dimofte:2014ija} and references therein.}. The 3d-3d correspondence is a rich relation between geometry of a 3-manifold $M_3$ and a 3d $\cN=2$ superconformal field theory $T_{\frakg}[M_3]$ arising from the partially twisted compactification of the 6d (2,0) theory with a simply-laced Lie algebra $\frakg$ on the 3-manifold $M_3$. In particular, the moduli space of flat $G_\bC$ connections on $M_3$ is identified with the moduli space of supersymmetric vacua in the  3d gauge theory  $T_{\frak g}[M]$ on $S^1\times D$. 
\bea\nonumber
\cM_{\textrm{SUSY}}(T_\frakg[M_3])\cong\cM_{\textrm{flat}}(M_3,G_\bC)
\eea
Moreover,  the partition function of $T_{\frak g}[M]$ on a lens space $L(k,1)$ is equal to the partition function of complex Chern-Simons theory with gauge group $G_\bC$  on $M_3$. 
\bea\nonumber
Z_{T_\frakg[M_3]}[L(k, 1)_b] = Z_{\textrm{CS}}^{(k,\sigma)}[M_3; G_\bC].
\eea
The level of $G_\bC$ complex Chern-Simons theory has a real part $k$ and an imaginary part $\sigma$, and $\sigma$ is related to the squashing parameter b of the lens space $L(k, 1)_b = S_b^3/\bZ_k$ by
\bea\nonumber
\sigma=k\frac{1-b^2}{1+b^2}~.
\eea
The physics derivation of the 3d/3d correspondence has been given by means of supersymmetric localization \cite{Yagi:2013fda,Lee:2013ida,Cordova:2013cea}. In particular, when a 3-manifold is a knot complement $M_3=S^3\backslash K$, using the 3d/3d correspondence and the Higgsing procedure, it was shown that the Poincar\'e polynomials of $\fraksl(2)$ homology of a knot $K$ can be realized as a partition function of the dual 3d $\cN=2$ theory $T_{\fraksl(2)}[S^3\backslash K]$ \cite{CDGS14}.

\vspace{.2cm}
\begin{center}
\begin{tikzpicture}[auto,node distance=2cm,
  thick,gauge node/.style={rounded rectangle,draw=blue!50,fill=blue!20},big node/.style={}]
    \node[big node] (11)  {};
   \node[gauge node] (21) at (0,-2) [scale=1] {3d $\mathcal{N}=2$ SCFT $T_\frakg[M_3]$ on $L(k,1)_b$};
  \node[gauge node] (12) at (4,0) [scale=1] {6d (2,0)  SCFT of $\frakg$ type   on $L(k,1)_b\times M_3$};
    \node [gauge node] (23) at (8,-2) [scale=1] {$G_{\bC}$ Chern-Simons  with level $(k,\sigma)$ on $M_3$};
    
\begin{scope}[->]
  \draw (12) to  node [left]{compactify on $M_3$  $\quad$ } (21);
    \draw (12) to  node [right]{$\quad$ localization on $L(k,1)_b$} (23); 
  \end{scope}
\end{tikzpicture}
\end{center}
\vspace{.2cm}

On the mathematics side, the program of formulating $\SL(N,\bC)$ Chern-Simons theory has just started \cite{Andersen:2014yna,Andersen:2014aoa}. In \cite{Andersen:2014yna}, the formulation of $\SL(2,\bC)$ quantum Chern-Simons theory  at level $k$ is given by means of the geometric quantization, which depends on the complex structure of an underlying surface. The independence of choice of complex structures is shown by constructing the projectively flat Hitchin-Witten connection. Moreover, using an ideal triangulations of a 3-manifold, mapping-class-group actions induced by the Hitchin-Witten connection are analyzed in \cite{Andersen:2014aoa}.

\subsection*{B-model and topological recursion}

The physical systems in \eqref{resolved} are realized as topological A-model so that it is natural to ask: what is the mirror manifold in the B-model? 
For non-compact toric Calabi-Yau $M$, the mirror manifold $ M^\vee$ \cite{Hori:2000ck}
is given by 
\be\nonumber
{M}^\vee=\{(u,v,x,y) \in\bC\times\bC\times\bC^*\times\bC^*| uv=H(x,y;a)\}
\ee
where the spectral  (holomorphic) curve $H(x,y;a)=0$ can be viewed as the moduli space of the canonical Lagrangian brane \cite{Aganagic:2000gs}. Here the mirror symmetry exchanges the K\"ahler parameter in the A-model with the complex structure $a$ of the spectral curve in the B-model. For instance, the spectral curve for the mirror  manifold of the resolved conifold \cite{Aganagic:2000gs} is expressed by
\bea\label{unknot-curve}
H_{\includegraphics[width=.3cm]{unknot}}(x,y;a)= 1-ax-y+a^{-1}xy~,
\eea
where  the canonical Lagrangian brane wraps the submanifold $\cL_{\includegraphics[width=.3cm]{unknot}}$ corresponding to the unknot.  Performing the  $\SL(2,\bZ)$ transformation of the curve \eqref{unknot-curve}, it was shown in  \cite{BEM,BEO13} that the configuration of the Lagrangian brane $\cL_{T_{m,n}}$ associated to the torus knot $T_{m,n}$ is encoded in the spectral curve 
\bea\label{BEM-torus}
H_{T_{m,n}}(x,y;a)= y^n(1-y)^m+a^{n-m}x(y-a^2)^m~.
\eea
Moreover, because the moduli space of a Lagrangian brane  $\cL_K$ receives the instanton corrections, even with the same resolved conifold background, the disc-corrected moduli space of $\cL_K$ depends on a knot $K$. Since the zero locus of the Q-deformed $A$-polynomial $A(K;x,y; a)$ determines the disc-corrected moduli space of $\cL_K$, it is argued  \cite{AV12} that the B-model geometry mirror to \eqref{resolved} is described by
\be\nonumber
uv=A(K;x,y; a)~.
\ee

Therefore, it is an important problem to understand how the mirror geometry encodes information about quantum knot invariants. Fortunately, there is a universal formalism, so-called \emph{topological recursion} \cite{Eynard:2007kz}, to calculate quantum and enumerative invariants based on the data of a spectral curve with two meromorphic functions $x$ and $y$, and a symmetric bi-differential $W_{0,2}$ on $\Sigma\times\Sigma$. In fact, the large color asymptotic expansion of colored Jones polynomials can be obtained by applying the topological recursion on the ordinary $A$-polynomial \cite{DFM10,BE12}. However, it is now known that the naive application of topological recursion on Q-deformed $A$-polynomial  does \emph{not} provide the large color asymptotic expansion of HOMFLY polynomials colored by symmetric representations. To circumvent this difficulty, a calibrated symmetric bi-differential (calibrated kernel) $W_{0,2}$ has been constructed in \cite{Gu} based on the data of the curve \eqref{BEM-torus}. With the calibrated symmetric bi-differential, the validity of the topological recursion on Q-deformed $A$-polynomial  is checked for torus knots.
Although a calibrated kernel is available only for a torus knot so far, it is expected that this is a technical restriction and the topological recursion based on the modified kernel works for more general knots as well.

\end{document}